\documentclass[a4paper,fleqn,usenatbib]{mnras}

\usepackage{newtxtext,newtxmath}

\usepackage[T1]{fontenc}
\usepackage{ae,aecompl}

\usepackage{graphicx}	
\usepackage{amsmath}	
\usepackage{amssymb}	
\usepackage{hyperref}
\usepackage{ragged2e}
\usepackage{multirow}
\usepackage{diagbox}

\title[Gas-dust correlations in NGC 3184 \& NGC 7793]{Gas-dust correlations in nearby galaxies: a case study of NGC 3184 and NGC 7793}

\author[Saikia, Patra, Roy \& Jog]{
Gautam Saikia,$^{1}$\thanks{E-mail (GS): gautamsaikia91@gmail.com }
Narendra Nath Patra,$^{2}$\thanks{E-mail (NNP): narendra@rri.res.in }
Nirupam Roy$^{1}$
and Chanda J. Jog$^{1}$
\\
\\
$^{1}$Department of Physics, Indian Institute of Science, Bengaluru 560012, India\\
$^{2}$Raman Research Institute, C. V. Raman Avenue, Bengaluru 560080, India\\
}

\pubyear{2019}
\date{}

\begin{document}
\label{firstpage}
\pagerange{\pageref{firstpage}--\pageref{lastpage}}
\maketitle

\begin{abstract}
The study of gas-dust interactions occurring in the interstellar medium of a galaxy is essential for understanding various physical processes taking place within it. A comparison of such events at different locations corresponding to diverse astrophysical environments provides more insight into the star-formation as well as dust destruction conditions and time-scales. We present a case study for two galaxies: NGC 3184 and NGC 7793, which are typical examples of a `grand design spiral' and a `flocculent spiral', respectively. We investigate the gas-dust correlations at various spatially resolved locations within each galaxy, including spiral arms, using archival data. Moreover, we have segregated the neutral gas into wide (warm) and narrow (cold) velocity components to check the correlations with individual dust emission bands. We find a positive correlation between the gas and the dust, with the total atomic gas emission mainly dominated by its warm component in both the galaxies. We also find the dust population in NGC 7793 to have a greater fraction of emission coming from cold and diffuse, larger-sized dust particles as compared to NGC 3184. This nearby galaxy pilot study could serve as a template for similar studies of larger galaxy samples with analogous morphologies.\\
\end{abstract}




\section{Introduction} 

The ubiquitous nature of dust particles in the universe and its interaction with gas in various environments play an important role in determining the characteristic emission/absorption seen from such regions. In dense interstellar gas or nebulae, which are considered as the birth-place of stars, dust may, in fact, dominate the atomic and molecular reactions which usually occur purely in the gas-phase. This can be seen spectroscopically along a particular line of sight, when there is an under-abundance of elements as compared to a Sun-like metallicity in the gas-phase, wherein the dust can be assumed to be the metal reservoir rather than the gas in the interstellar medium (ISM) \citep{Dyson+1997}. It is now well known that the incoming ultraviolet (UV) and visual radiation from stars are absorbed and re-radiated by dust particles, mainly in a continuum, as heat in the infrared (IR) which allows the dust to maintain a grain temperature $\sim$30 K. They also scatter the shorter wavelength light and a combination of the two processes is known as extinction \citep{Trumpler+1930}. The composition of these dust particles is known from the type of emission/absorption spectra they produce with elemental peaks/troughs overlaying the dust continuum.\\

Cosmic dust is usually found to be mixed with the dense (n$_{H}$>10$^{2}$ cm$^{-3}$) and cool (T <$\sim$10$^{2}$ K) interstellar gas clouds, which account for a majority of the mass in the ISM despite being smaller in volume. This dust not only aids in regulating the heating/cooling of the ISM by scattering/absorbing the incident ultraviolet radiation and re-emitting in the infrared \citep{Draine+2011}, but also provides a site for molecular formation to occur (including H$_2$) and is an important agent in providing an ionizing balance in the ISM, with a major role in the star formation process \citep{Draine+2003}. The ionized gas nebulae, also known as HII regions, have hot stars embedded inside them which maintain a high temperature ($\sim$10$^4$ K), with a density (>10$^{2}$ cm$^{-3}$) similar to the neutral regions surrounding it. Dust particles generally do not survive such harsh environments and are usually found in neutral molecular clouds or in intercloud gas regions (usually atomic) which have been partly ionized by X-rays from hot stars making them quite warm ($\sim$10$^{3}$ K) with low density (< 1 H atom cm$^{-3}$) \citep{Williams+2005}. The emission by dust particles and its temperature distribution depends on several factors including the grain size, composition, opacity and strength of the local interstellar radiation field. Hence, given the important role played by dust in the universe around us, the study of gas-dust interactions becomes highly crucial to understanding the underlying physics of the ISM and high-density star-forming regions in various astrophysical systems.\\

Surveys such as the Spitzer Infrared Nearby Galaxy Survey (SINGS, \cite{Kennicutt+2003}) and the GALEX Nearby Galaxy Survey (GANGS, \cite{GildePaz+2007}) in the IR and UV respectively, have made enormous contributions to understanding the star formation processes occurring in nearby galaxies. Such studies also require a reservoir of molecular gas data products, in addition to the star and dust data, which have been made available at complementary resolutions thanks to the BIMA Survey of Nearby Galaxies (BIMA SONG, \cite{Helfer+2003}) and the HI Nearby Galaxy Survey (THINGS, \cite{Walter+2008}) corresponding to CO and HI surveys, respectively. In the last decade, the Heterodyne Receiver Array CO Line Extragalactic Survey (HERACLES, \cite{Leroy+2009}) and the Key Insights on Nearby Galaxies: a Far-Infrared Survey with \textit{Hershel} (KINGFISH, \cite{Kennicutt+2011}) surveys have made high-resolution CO and far-IR (FIR) maps available which have allowed us to trace the gas and dust mass distribution in nearby galaxies at $\sim$kiloparsec (kpc) scales, which extend into the HI-dominated ISM.\\

\cite{Leroy+2008}, \cite{Bigiel+2008} and \cite{Leroy+2012} have made use of various data sets from these high-resolution surveys including the THINGS, SINGS, BIMA SONG and HERACLES to investigate the relationship between star-formation rates (SFRs) and neutral gas distribution/density in nearby galaxies at sub-kpc scales. \cite{Leroy+2008} found that depending on the type of galaxy (spiral/irregular) and location (inner/outer disk), the same gas surface density can lead to very different (SFRs), which suggests that local effects such as potential well, chemical enrichment, pressure and coriolis forces/shear may control how star formation occurs in the neutral phase of the ISM. \cite{Leroy+2012} have tried to estimate the contamination of emission observed at 24 $\mu$m, not associated with recent star formation. Among others, \cite{Foyle+2010} have studied the star formation rates and how they vary in the arm and inter-arm regions for a sample of spiral galaxies; and found a significant contribution (at least 30\%) of the inter-arm regions to SFR tracers. Using FIR observations from KINGFISH, CO from HERACLES and HI from THINGS, \cite{Sandstrom+2013} present the CO-to-H$_2$ conversion factor ($\alpha_{CO}$) and the dust-to-gas ratio (DGR), at $\sim$kpc resolution, in a sample of nearby galaxies. A close empirical correlation has been found between the FIR luminosity, which is known as a young star formation tracer, and the synchrotron-dominated centimetre-wavelength emission \citep{Helou+1985}; which is why the radio continuum emission is frequently used as a dust-free tracer of recent star formation in galaxies. Observations made by the \textit{Spitzer} telescope found that the FIR 70 $\mu$m and non-thermal radio emission correlation seems to show a dependence on the time-scale of recent star formation occurrence as the cosmic ray electrons in such cases do not get enough time to diffuse over significant distances, which is termed as ``age effect" \citep{Murphy+2008}. While high energy stellar UV photons from OB stars are expected to dominate the heating of gas near star-forming regions, it could also be due to turbulence, shocks and collisions among clouds in more isolated environments \citep{Flower+2010}. Hence, there arises a need to study the underlying radio-IR correlation mechanism using recently available high-resolution data \citep{Kennicutt+2011}.\\

In this work, we report a systematic study of the gas-dust correlations for two nearby spiral galaxies: NGC 3184 and NGC 7793. We have separated the wide and narrow components of the neutral gas and investigated the correlations independently for multiple dust emission bands. These correlations eventually give us an idea of the gas-dust interactions, which we have compared for the two galaxies. We present the first such resolved gas-dust study with linear resolution $\sim$few hundred parsecs using archival data. A brief description of our galaxy sample is presented in Section \ref{sample}. The procedure for the analysis of archival data is presented in Section \ref{analysis}. The results of correlation studies and their interpretations are presented in Section \ref{results} followed by the main conclusions of this work in Section \ref{conclusions}. \\ 
 
\section{Target selection \& description}
\label{sample}

\subsection{Galaxy sample}

The aim of this study is to look for a pixel-by-pixel correlation between gas and dust emissions and hence, our sample must have data available in both the radio and infrared (IR). In the near and mid-IR, beam sizes are small and hence the spatial resolution of instruments is not much of a problem. However, this is usually not the case for single-dish radio observations which is why we select a sample of galaxies such that it is sufficiently nearby, in order to avoid any contamination in the observed spectra due to rotation effects. We have selected NGC 3184 and NGC 7793 which are classified as a `grand design spiral' and a `flocculent spiral', respectively, according to the scheme of morphological galaxy classification given by \cite{Elmegreen+1982}. The difference in properties such as star-formation rate, inclination angle, etc. make these two ideal for a study of the trend in gas-dust correlations at high linear resolution among galaxies having very different spiral structures, but closely related in morphological type (Table \ref{table:galaxies}). The findings reported here are part of an ongoing larger project making use of archival data for nearby galaxies and we present these two example galaxies as a pilot study. \\

\begin{figure*}
\centering
\includegraphics[width=8cm,height=6cm]{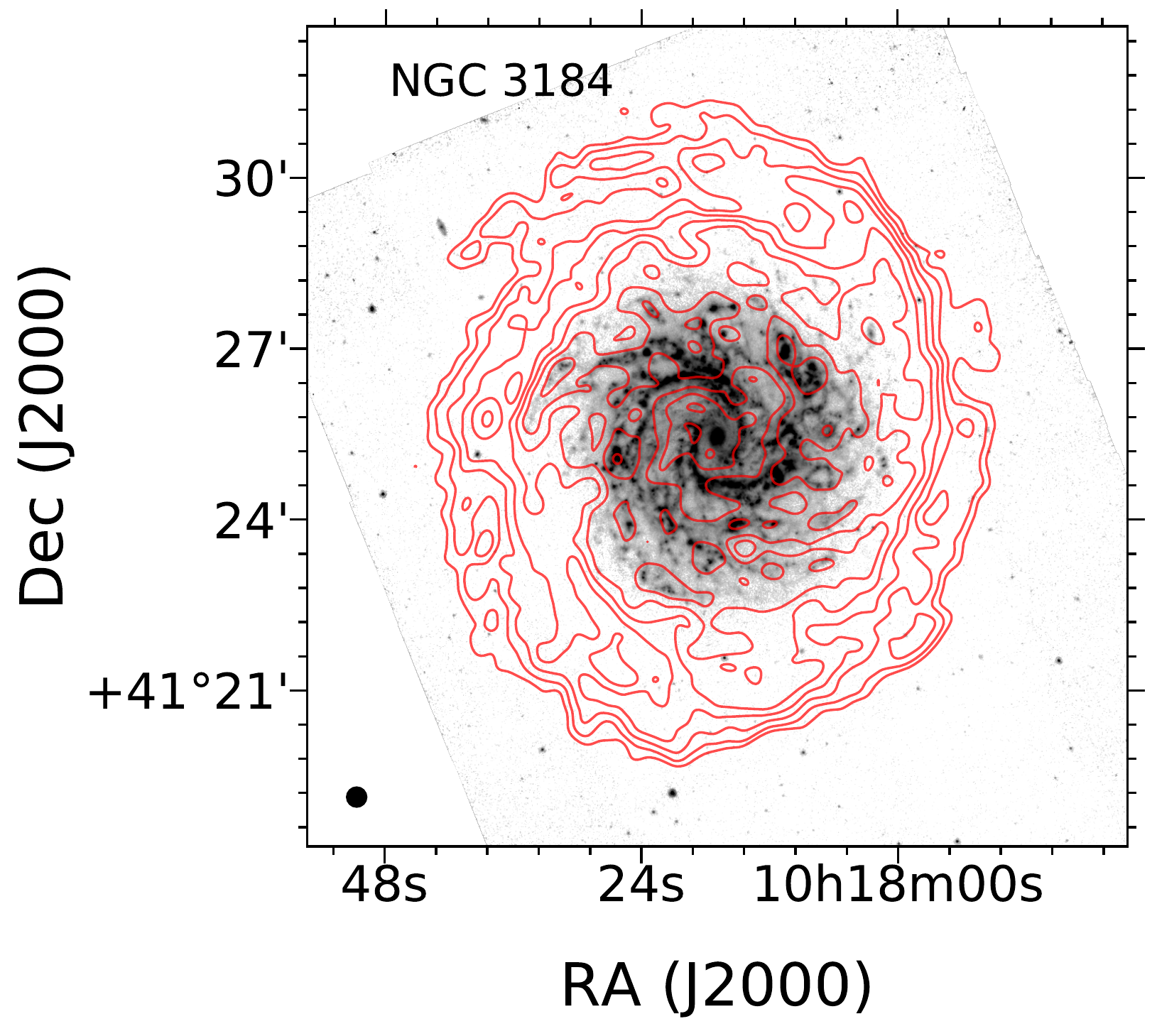}
\includegraphics[width=8cm,height=6cm]{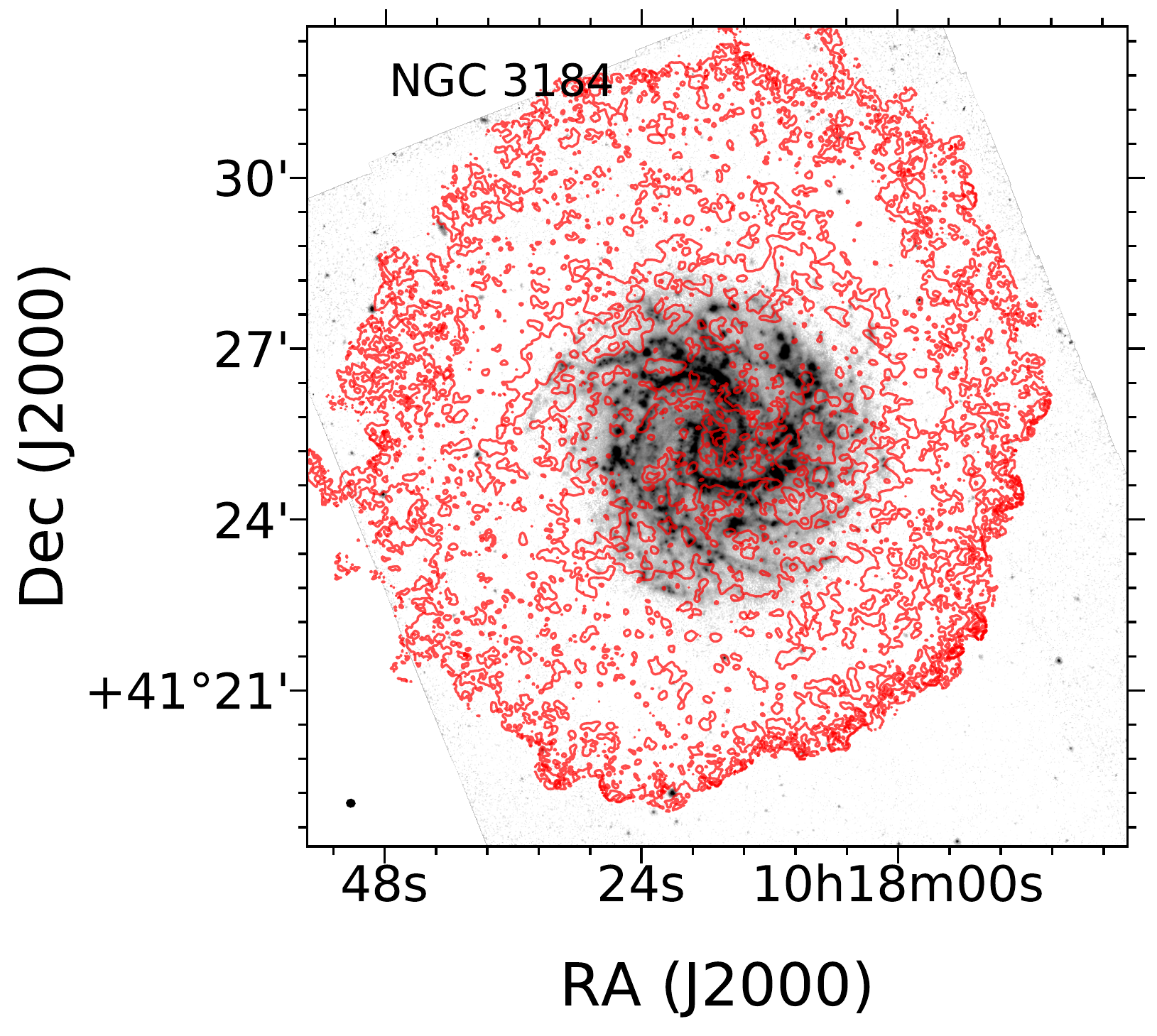}
\caption{SINGS 8 $\mu$m dust images of NGC 3184 (in grey) overlaid with THINGS moment 0 (MOM0) integrated HI intensity (left panel) and moment 2 (MOM2) velocity dispersion (right panel) maps, shown as contours (in red). The MOM0 contour levels are (1, 1.4, 2, 2.8, ...)$\times$10$^{20}$ atoms/cm$^{2}$ and MOM2 contour levels are (5, 10, 15, 20, ...) km/sec. The respective HI beam sizes are shown as a black dot in the bottom left corner of each panel. The well-defined spiral arms of the `grand design spiral' NGC 3184 are clearly visible in both panels.}
\label{overlays_3184}
\end{figure*}

\begin{figure*}
\centering
\includegraphics[width=8cm,height=6cm]{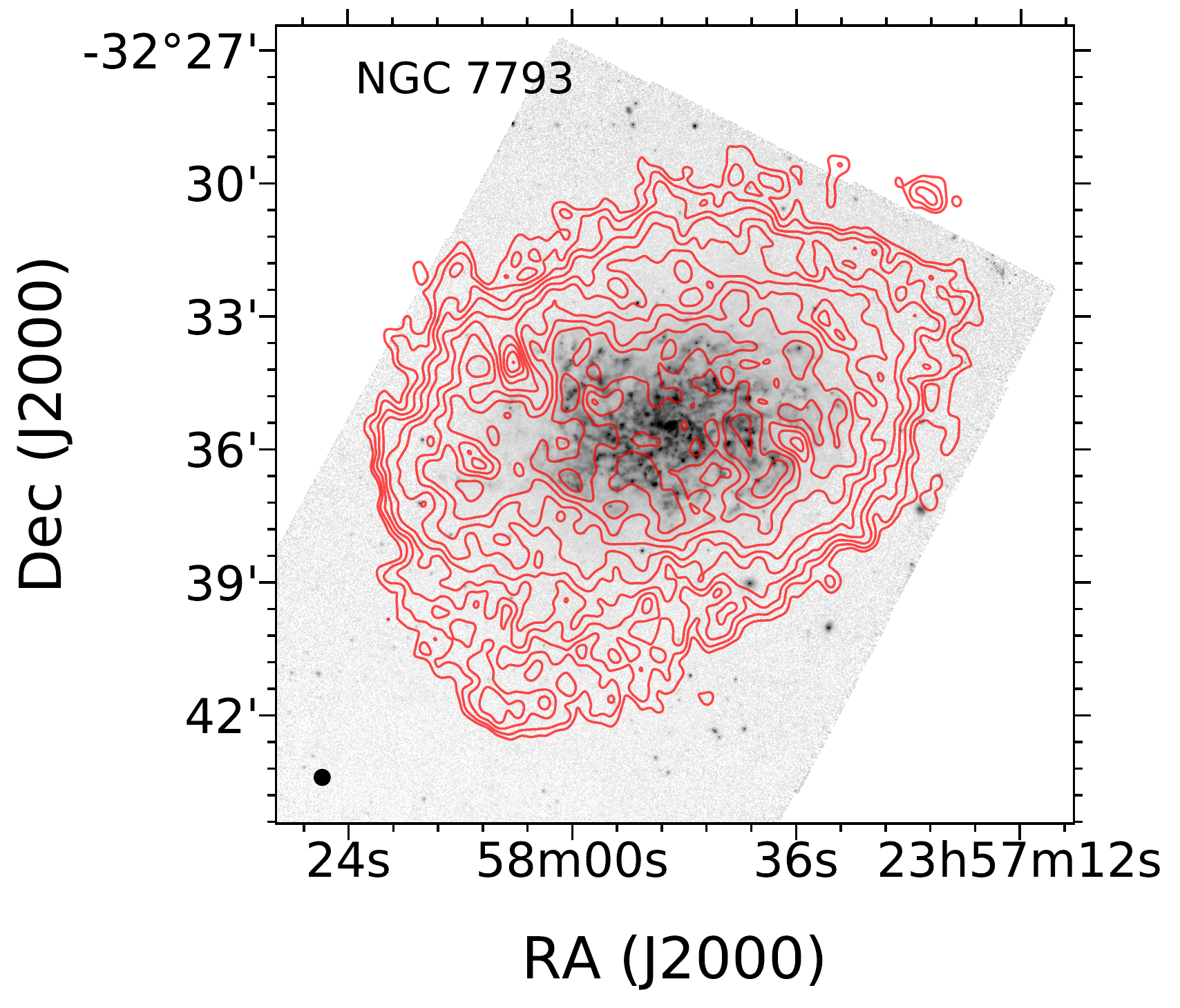}
\includegraphics[width=8cm,height=6cm]{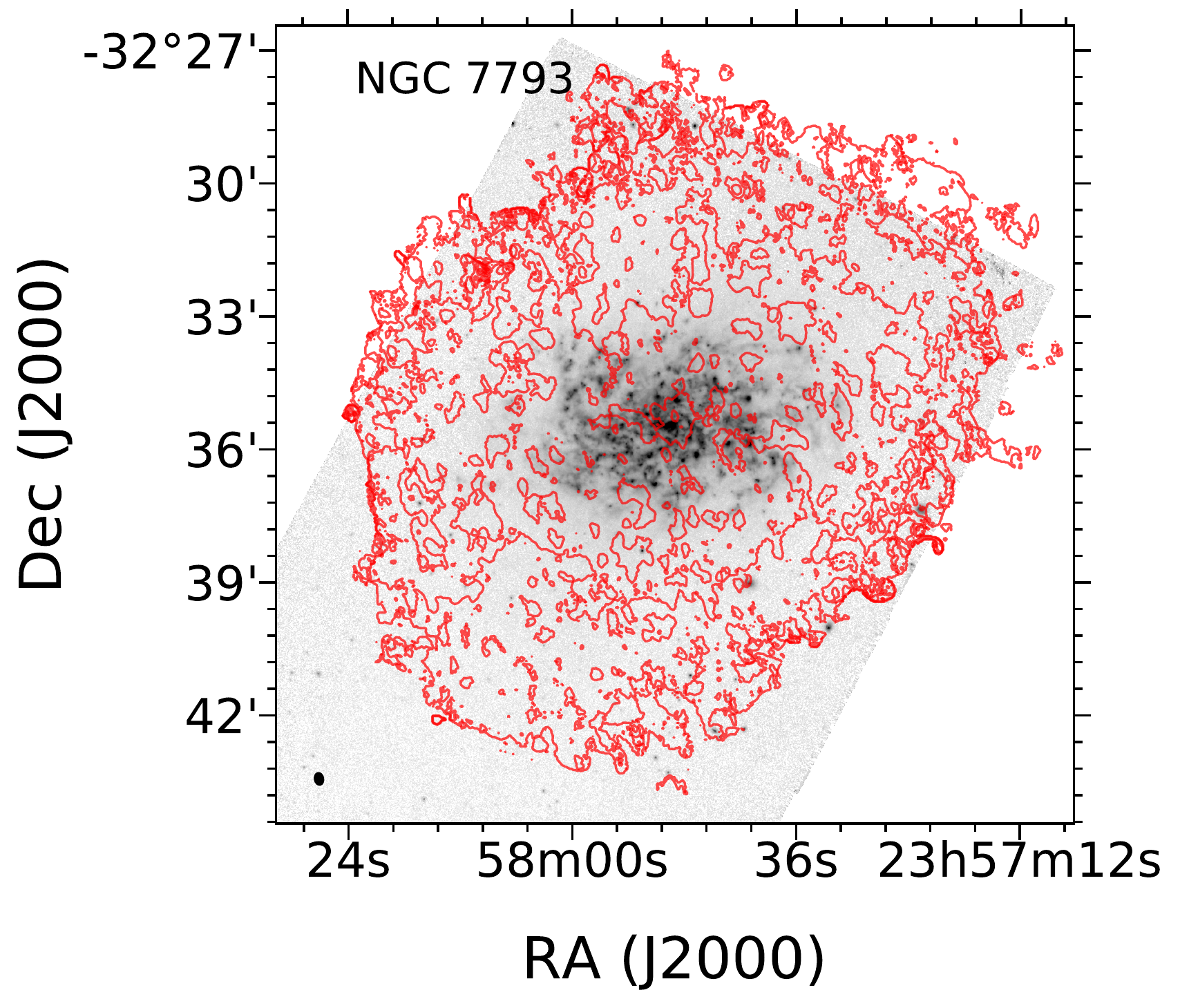}
\caption{SINGS 8 $\mu$m dust images of NGC 7793 (in grey) overlaid with THINGS moment 0 (MOM0) integrated HI intensity (left panel) and moment 2 (MOM2) velocity dispersion (right panel) maps, shown as contours (in red). The MOM0 contour levels are (1, 1.4, 2, 2.8, ...)$\times$10$^{20}$ atoms/cm$^{2}$ and MOM2 contour levels are (5, 10, 15, 20, ...) km/sec. The respective HI beam sizes are shown as a black dot in the bottom left corner of each panel. The `flocculent spiral' NGC 7793 can be distinctly differentiated from NGC 3184 due to a lack of definite spiral arms.}
\label{contours_7793}
\end{figure*}

\begin{table*}
\scriptsize 
\centering
\caption{A compilation of some known properties for the two galaxies: NGC 3184 and NGC 7793 in our sample.}
\label{table:galaxies}
\begin{tabular}{cccccccccccc}
\hline
\hline
Galaxy & RA (J2000)$^{a}$ & DEC (J2000)$^{a}$ & Morph.$^{a}$ & Size$^{a}$ & Dist.$^{a}$ & Incl.$^{b}$ & log[M(HI)]$^{c}$ & log[M(H$_2$)]$^{c}$ & log(M$_{dust}$)$^{c}$ & L$_{TIR}^{d}$ \\
     & (hh mm ss) & (dd mm ss) &  & (arcmin) & (Mpc) & (degrees) & (M$_{\odot}$) & (M$_{\odot}$) & (M$_{\odot}$) & (L$_{\odot}$) \\
\hline
NGC 3184 & 10 18 16.98 & +41 25 27.77 & Scd & 7.4$\times$6.9$^{\prime}$ & 11.3 & 16 & 9.26 & 9.11 & 7.70 & 1.1$\times$10$^{10}$ \\
NGC 7793 & 23 57 49.75 & -32 35 27.71 & Sd & 9.3$\times$6.3$^{\prime}$ & 3.6 & 50 & 8.77 & -- & 6.92 & 2.3$\times$10$^{9}$ \\
\hline
\hline
\end{tabular}
\begin{flushleft}
\footnotesize
\textbf{Notes:}\\
$^{a}$ Coordinates, morphological types, optical sizes and distances as listed in NED and SIMBAD astronomical databases.\\
$^{b}$ Inclination angles from \cite{Walter+2008}. \\
$^{c}$ Atomic [M(HI)], molecular [M(H$_2$)] gas and dust (M$_{dust}$) masses from \cite{Draine+2007SINGS}.\\
$^{d}$ Total infrared luminosity (TIR) in the 3-1000 $\mu$m range from \cite{Kennicutt+2011}.\\
\end{flushleft}
\end{table*}

\vspace{1cm}

\begin{center}
\textbf{NGC 3184}\\
\end{center}
NGC 3184 has strong and definite spiral arms such that it is classified as a ``grand design spiral", as seen in  Figure \ref{overlays_3184}. Located at a distance of 11.3 Mpc, it has a nearly face-on orientation (i=16 deg) which makes it an ideal candidate to study gas-dust interactions in the galaxy. Being a galaxy similar to the Milky Way, NGC 3184 has been used as a sample galaxy to determine the spatial correlation between diffuse hot gas and star-forming regions by \cite{Doane+2004}. They observed a correlation between regions of diffuse X-ray emission and the spiral arms and HII regions, and also found the diffuse thermal emission to be widely spread out across NGC 3184 with an average surface brightness similar to what is seen in the Milky Way neighbourhood. There have been multiple attempts to determine parameters such as the star-forming efficiency, CO-to-H$_{2}$ conversion factor ($\alpha_{CO}$), dust-to-gas ratio (DGR) and metallicity within this galaxy. \cite{Sandstrom+2013} found that the DGR was almost linearly correlated while the $\alpha_{CO}$ was only weakly-correlated with the metallicity in NGC 3184. \cite{Honig+2015} have determined the positions of 180 HII regions distributed along the two prominent spiral arms in NGC 3184. More recently, \cite{Abdullah+2017} have tried to quantify the contribution of various ISM phases in NGC 3184 to the [CII] emission, which is used as a star formation rate indicator. \\

\begin{center}
\textbf{NGC 7793}\\
\end{center}
NGC 7793 falls into the ``flocculent spiral" category \citep{Elmegreen+1982} which means `fluffy or wool-like', characterized by diffuse, broken spiral arms, with no bar and a very faint central bulge as shown in Figure \ref{contours_7793}. Such spiral structures were claimed to have been formed due to stochastic self-propagating star formation (SSPSF model; \citet{Mueller+1976}). Subsequently, these features have been attributed to being material arms arising due to sheared star-gas gravitational instabilities \citep{Jog+1984, Sellwood+1984, Jog+1992}. At a distance of only 3.6 Mpc and an apparent diameter of 10$^\prime$, NGC 7793 forms a part of the Sculptor Group, which is one of the closest groups of galaxies to the Local Group. It is neither an edge-on nor a completely face-on galaxy (i=50 deg) composed of highly ionized gas and has a mass, size and morphology similar to M33, which has prompted it to be a primary candidate for the study of galaxy kinematics in recent years. \cite{Muraoka+2016} have investigated the relationship between CO and total IR luminosities in NGC 7793 and found a linear correlation between them which is consistent even for ultraluminous infrared and submillimeter galaxies. They also found a linear correlation between the CO intensities and star-formation rates which they suggest is universally applicable to various types and spatial scales of galaxies.\\

A compilation of some known properties of the two galaxies is presented in Table \ref{table:galaxies}. Historically, NGC 3184 has been the more widely studied galaxy of the two. However, a study of gas-dust correlations by separating the warm and cold components of HI gas has not been attempted before in either case. Hence, while both galaxies are individually interesting due to the similarities they possess with the Milky Way and M33, respectively, we wanted to study how the gas-dust interactions vary within each galaxy with a change in physical properties. We have, in fact, separated the spiral arm locations to further investigate the possible contribution of local effects in NGC 3184.\\

\subsection{Radio HI data}

The HI Nearby Galaxy Survey (\href{http://www.mpia.de/THINGS/Data.html}{THINGS}), carried out using the NRAO Very Large Array (VLA), was an attempt to probe the atomic interstellar medium of 34 nearby galaxies at high spectral ($\leq$5.2 km/s) and spatial ($\sim$6$^{\prime\prime}$) resolution \citep{Walter+2008}. The galaxy sample was selected such that a wide range of physical properties, metallicities, star-formation rates and absolute luminosities were covered at 2$\leq$D$\leq$15 Mpc distances ($\sim$100-500 pc linear resolution). The high angular resolution allowed the observer to trace the neutral HI complexes and to resolve spiral arms. For each galaxy, HI data cubes as well as moment maps (MOM0: integrated HI maps, MOM1: mean intensity-weighted velocity, MOM2: velocity dispersion), are publicly available. Both NGC 3184 and NGC 7793 form a part of the THINGS galaxy sample.\\ 

\subsection{Infrared data}

Most of the THINGS galaxies were chosen to overlap with the \textit{Spitzer} Infrared Nearby Galaxies Survey (\href{https://irsa.ipac.caltech.edu/data/SPITZER/SINGS/summary.html}{SINGS}) \citep{Kennicutt+2003} sample, which was a mid and far-infrared emission survey of 75 nearby galaxies (D$<$30 Mpc). The SINGS survey aimed at a better understanding of how star formation is affected by ISM properties. Each of these galaxies was mapped by four IRAC (Infrared Array Camera; \citet{Fazio+2004}): 3.6, 4.5, 5.8, 8 $\mu$m and three MIPS (Multiband Imaging Photometer for \textit{Spitzer}; \citet{Rieke+2004}): 24, 70, 160 $\mu$m bands. The THINGS spatial resolution is complementary to that of MIPS 24 $\mu$m ($\sim$5.7$^{\prime\prime}$) which makes them very useful datasets for comparative studies. Moreover, the 8 $\mu$m mid-IR (MIR) band is generally attributed to polycyclic aromatic hydrocarbon (PAH) emission while the 24 $\mu$m emission is seen from very small dust grains near hot star-forming regions \citep{Draine+2003}. While 8 $\mu$m emission highlights the rims of HII regions, the 24 $\mu$m emission comes from mainly within the HII region where PAHs cannot survive. This has prompted their use as tracers of star-formation \citep{Dale+2005,Calzetti+2007}. However, the resolution of the MIPS 70 and 160 $\mu$m bands is quite poor as compared to the MIR (17$^{\prime\prime}$ and 38$^{\prime\prime}$ respectively; corresponding to 1--5 kpc linear scales), which led us to look for alternative observations to trace dust emission in the FIR.\\

\begin{table*}
\centering
\caption{A comparative chart showing details of imaging observations using different telescopes/instruments available for the two galaxies. The molecular CO data for NGC 7793 was unavailable, as per our requirements.}
\label{table:telescopes}
\begin{tabular}{ccccc}
\hline
\hline
Survey & Instrument & Observation & Beam size/PSF & Pixel size\\
\hline
THINGS & NGC 3184/ VLA & HI 21 cm  & 7.5$^{\prime\prime}\times$6.9$^{\prime\prime}$ & 1.5$^{\prime\prime}$ \\
       & NGC 7793/ VLA & HI 21 cm  & 15.6$^{\prime\prime}\times$10.8$^{\prime\prime}$ & 1.5$^{\prime\prime}$ \\
\hline
SINGS & IRAC/ \textit{Spitzer} & 8 $\mu$m & 2.0$^{\prime\prime}\times$2.0$^{\prime\prime}$ & 0.7$^{\prime\prime}$ \\ 
 & MIPS/ \textit{Spitzer} & 24 $\mu$m & 5.7$^{\prime\prime}\times$5.7$^{\prime\prime}$ & 1.5$^{\prime\prime}$ \\ 
\hline
 & PACS/ \textit{Hershel} & 70 $\mu$m & 5.8$^{\prime\prime}\times$5.5$^{\prime\prime}$ & 1.4$^{\prime\prime}$ \\
KINGFISH & PACS/ \textit{Hershel} & 100 $\mu$m & 6.9$^{\prime\prime}\times$6.7$^{\prime\prime}$ & 1.7$^{\prime\prime}$ \\
 & PACS/ \textit{Hershel} & 160 $\mu$m & 12.1$^{\prime\prime}\times$10.6$^{\prime\prime}$ & 2.8$^{\prime\prime}$ \\
\hline
HERACLES & NGC 3184/ IRAM-30m & CO (J=2$\rightarrow$1) & 13.4$^{\prime\prime}\times$13.4$^{\prime\prime}$ & 2.0$^{\prime\prime}$ \\
\hline
\hline
\end{tabular}
\end{table*}

The Key Insights on Nearby Galaxies: a Far-Infrared Survey with \textit{Hershel} (\href{https://irsa.ipac.caltech.edu/data/Herschel/KINGFISH/summary.html}{KINGFISH}) project \citep{Kennicutt+2011} had overlapping goals with the SINGS survey but covered exclusively the FIR at a much better resolution, which works to our advantage. The \textit{Hershel} PACS (Photodetector Array Camera and Spectrometer; \citet{Poglitsch+2010}) observed all 61 galaxies in the KINGFISH sample at 70, 100 and 160 $\mu$m with a fourfold improvement in spatial resolution over \textit{Spitzer} MIPS at similar wavelengths. This allows us to trace the increase in contribution from diffuse dust emission (cirrus) with increasing wavelength \citep{Bendo+2006} at an excellent spatial resolution for both NGC 3184 and NGC 7793.\\

\subsection{CO data}

The complementary CO data for NGC 3184 were available from both the Berkeley Illinois Maryland Association Survey of Nearby Galaxies (\href{https://ned.ipac.caltech.edu/level5/March02/SONG/SONG.html}{BIMA SONG}, \cite{Helfer+2003}) and the Heterodyne Receiver Array CO Line Extragalactic Survey (HERACLES, \cite{Leroy+2009}) archives, with different resolutions. While the BIMA SONG survey used a single-dish technique to systematically image the CO(J=1$\rightarrow$0) molecular emission at 3 mm with a typical resolution of 6$^{\prime\prime}$ or 360 pc at an average of 12 Mpc distance for the galaxy sample, the HERACLES survey used the single-dish 30-m IRAM telescope to map the CO(J=2$\rightarrow$1) line at 13$^{\prime\prime}$ resolution ($\sim$500 pc). Although the BIMA SONG and HERACLES surveys trace different CO transitions, \cite{Braine+1993} had observed both the emissions for a sample of 81 galaxies and found a typical line ratio of CO (J=2$\rightarrow$1/1$\rightarrow$0) of 0.89$\pm$0.06, which supports the use of CO(J=2$\rightarrow$1) emission as a viable molecular hydrogen (H$_2$) tracer in other galaxies. Moreover, the CO(J=2$\rightarrow$1) follows the same distribution as that of CO(J=1$\rightarrow$0) in our Milky Way \citep{Israel+1984,Sakamoto+1995}. The main advantage of using a single dish in comparison to an interferometer for such observations lies in the sensitivity of such an arrangement to extended structures such as molecular cloud complexes which may be missed by high resolution, low sensitivity interferometric observations. We have used the HERACLES data for NGC 3184 in this study which can adequately resolve spiral arms, bars as well as large scale star-forming complexes \citep{Leroy+2009}. The good extent and sensitivity of these maps, which often measured the CO beyond the transition zone of H$_2$-to-HI, made it possible to clearly distinguish between the total gas and its molecular component \citep{Bigiel+2008}. The CO data for NGC 7793 was unavailable in both BIMA SONG and HERACLES archives. \\

A comparative chart of the resolutions at different wavelengths for the four surveys (THINGS, SINGS, KINGFISH, HERACLES) is presented in Table \ref{table:telescopes}. An overlay of the THINGS radio HI intensity and velocity dispersion contours on an 8 $\mu$m SINGS image is shown in Figure \ref{overlays_3184} for NGC 3184 and Figure \ref{contours_7793} for NGC 7793.\\

\section{Data analysis}
\label{analysis}

\subsection{Two-component HI maps}

\begin{figure*}
\centering
\includegraphics[width=6.5cm,height=5.5cm]{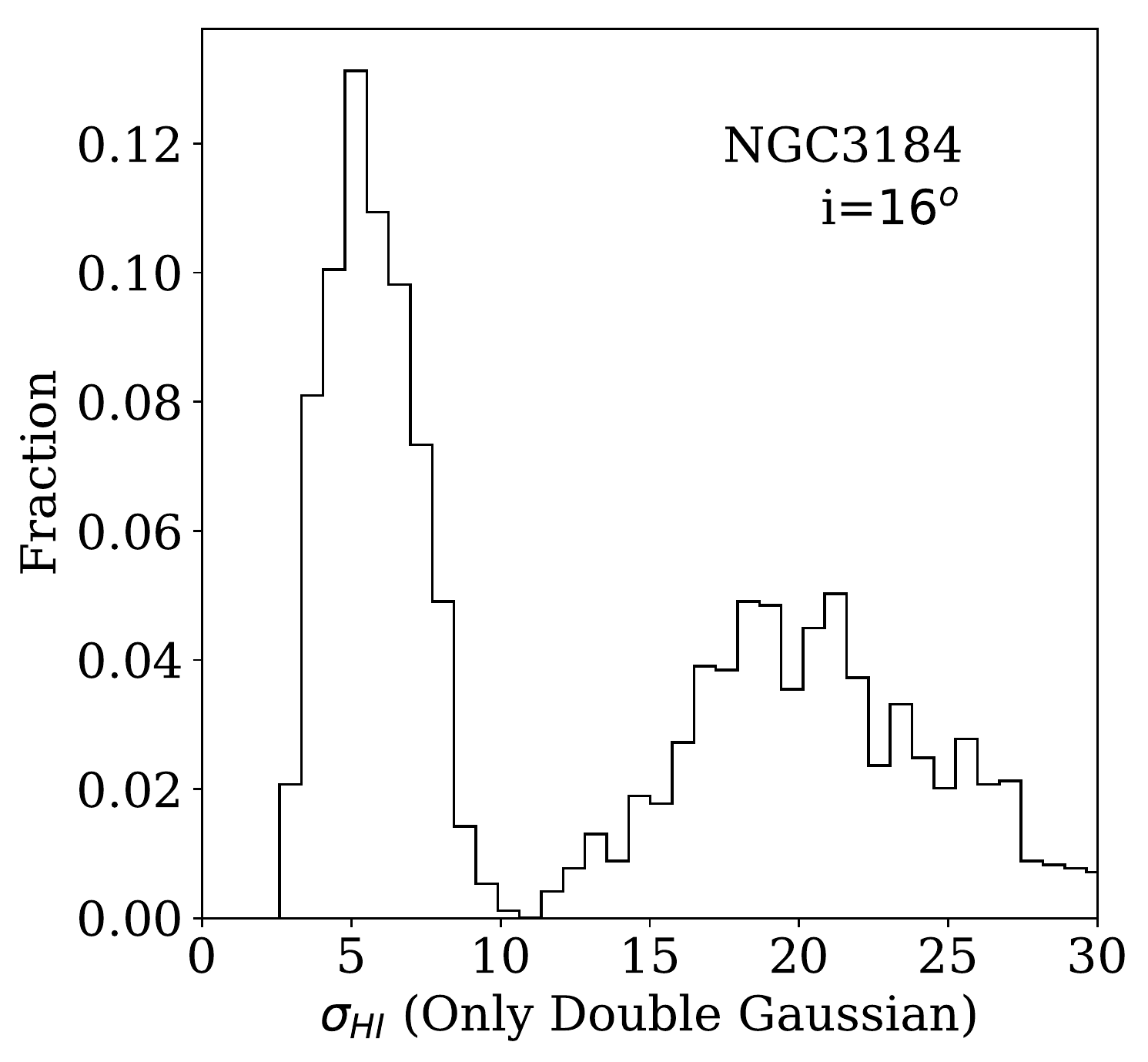}
\includegraphics[width=6.5cm,height=5.5cm]{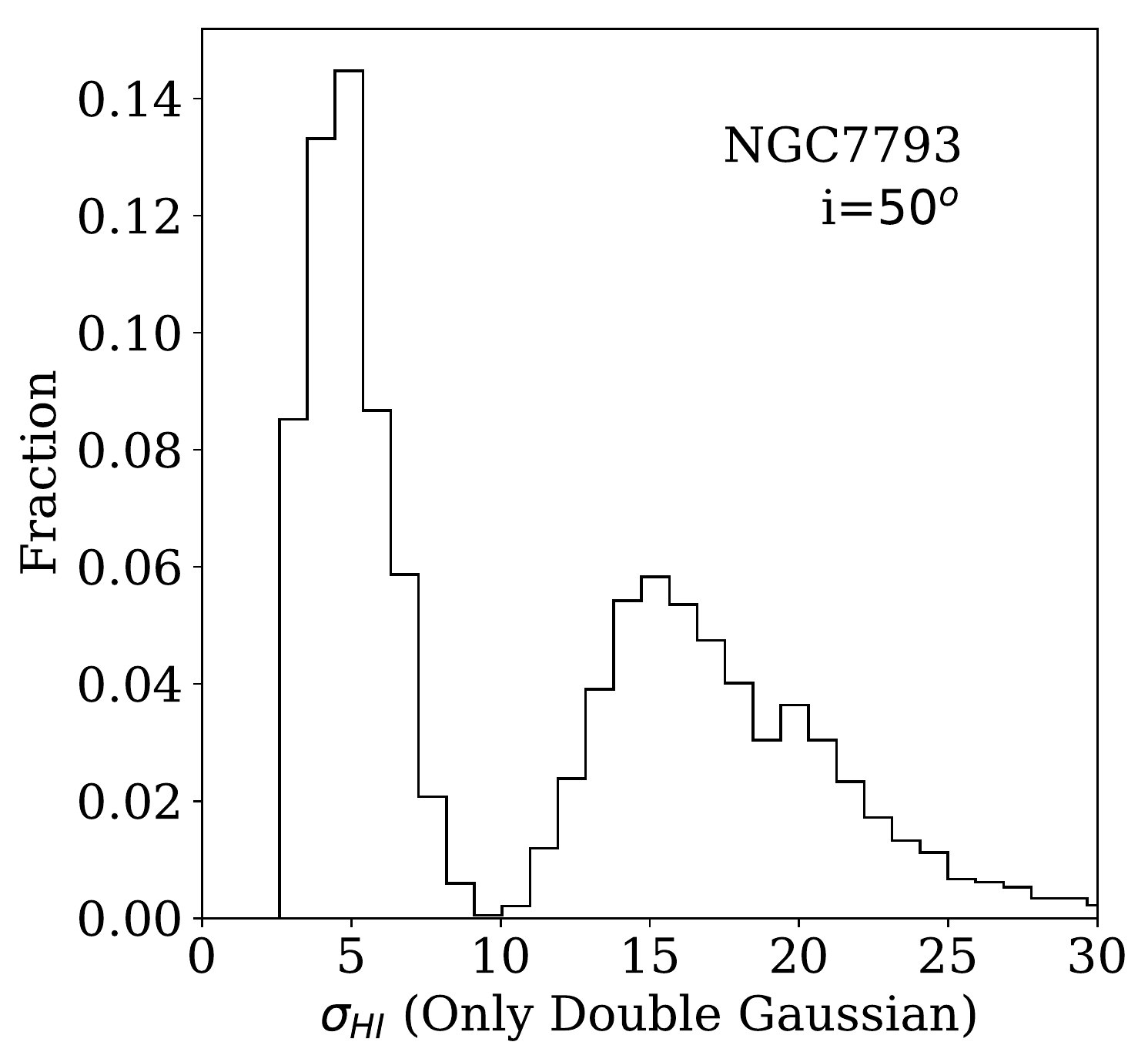}
\caption{The histograms of the decomposed $\sigma_{HI}$ values for NGC 3184 (left panel) and NGC 7793 (right panel) achieved by fitting a double Gaussian to each pixel. The threshold for separation of narrow (cold) and wide (warm) components at $\sigma_{HI}$=10 km/s is clearly seen in both panels.}
\label{HImap_sig}
\end{figure*}

We have used the procedure detailed in \cite{Patra+2016} to separate the wide and the narrow velocity components of the atomic HI gas for each galaxy, in order to individually study their effects on the dust emission in the region.  Each line-of-sight spectrum of the galaxy is fit first by a single and then a double Gaussian. The residuals are compared for both cases and an F-test (statistical test) is done to find out if adding an extra Gaussian component (i.e., double Gaussian) does improve the fit with reasonable statistical significance (95\%) or not. We did this exercise to every pixel (with an automated routine) and then extracted the corresponding velocity dispersion ($\sigma_{HI}$) values. Hence, in the total data cube, there are some pixels which are best fitted by single Gaussian and there are some which are best fitted by double Gaussian giving us two histograms per galaxy: (1) $\sigma_{HI}$ values where the best fit to the HI spectra was achieved by fitting a single Gaussian, (2) spectra best fitted by a double Gaussian. In the case of single Gaussian fit, there is no clear demarcation of wide and narrow components ($\sigma_{HI}$) in either galaxy and it looks more like a continuous distribution. On the other hand, we clearly see two distinct distributions in the histograms for the double Gaussian ($\sigma_{HI}$) fit for both galaxies (Figure \ref{HImap_sig}). \\

\begin{figure*}
\centering
\includegraphics[width=8cm,height=6cm]{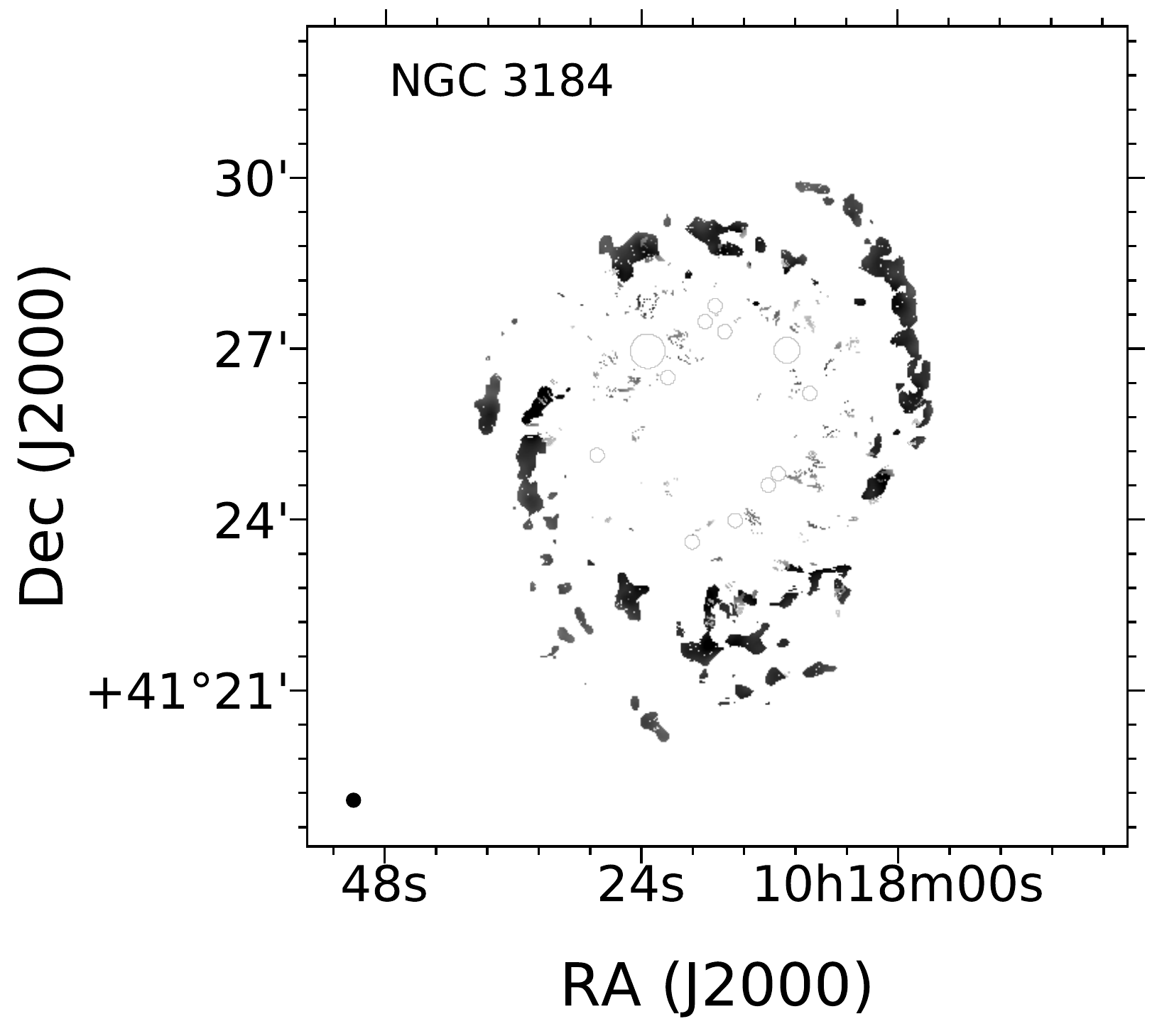}
\includegraphics[width=8cm,height=6cm]{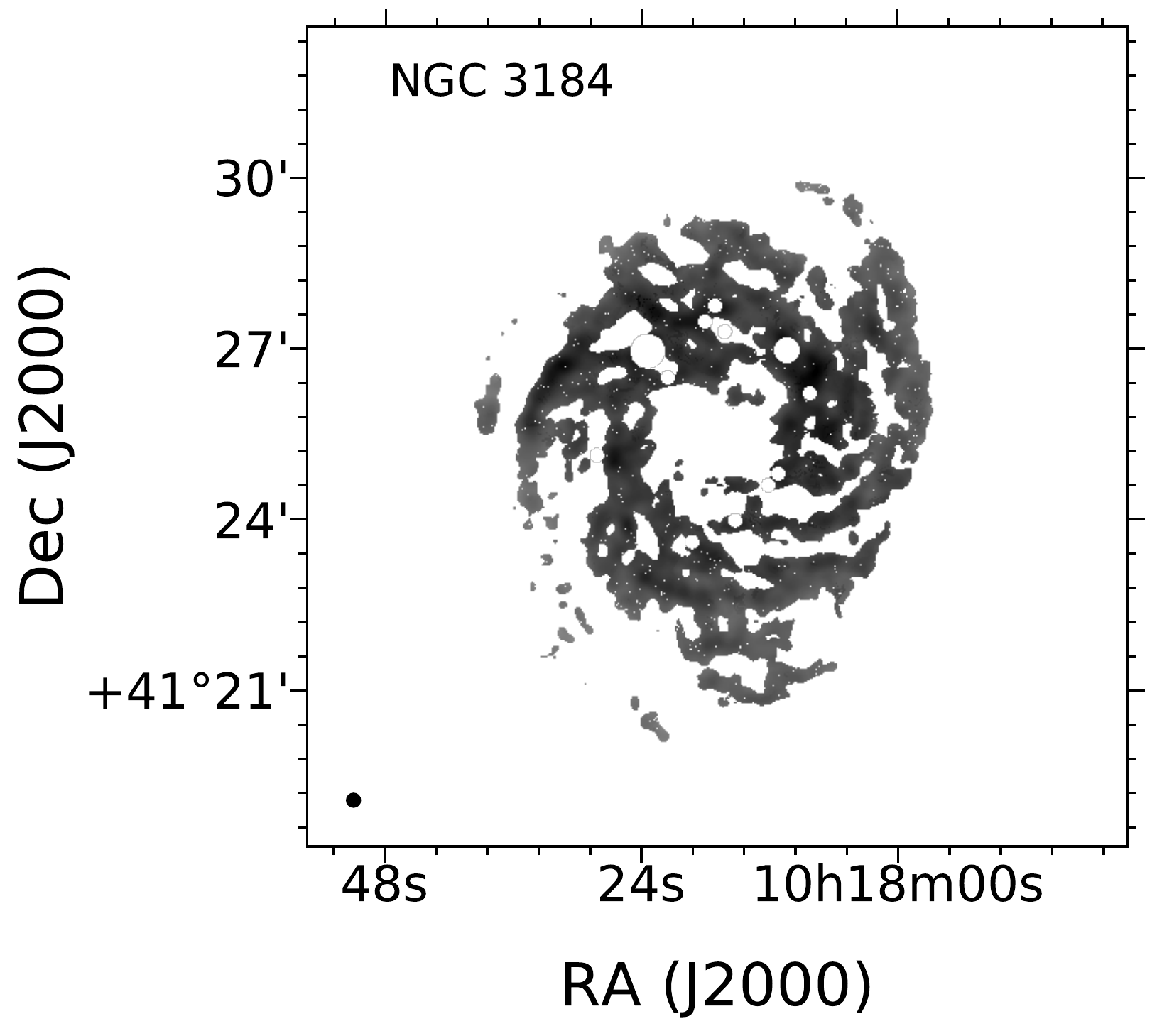}
\caption{The cold (left panel) and warm (right panel) components of HI gas separated into two images using the method described in \citet{Patra+2016}. We see a distinct bi-modality of the atomic gas distribution in NGC 3184. The masks applied to avoid contamination from Galactic foreground stars can be seen as white circles with the convolved beam size (13.4$^{\prime\prime}\times$13.4$^{\prime\prime}$) shown as a black dot in the bottom left corner of each panel. The angular size of these images has been kept the same as Figure \ref{overlays_3184} for comparison.}
\label{3HImaps}
\end{figure*}

\begin{figure*}
\centering
\includegraphics[width=8cm,height=6cm]{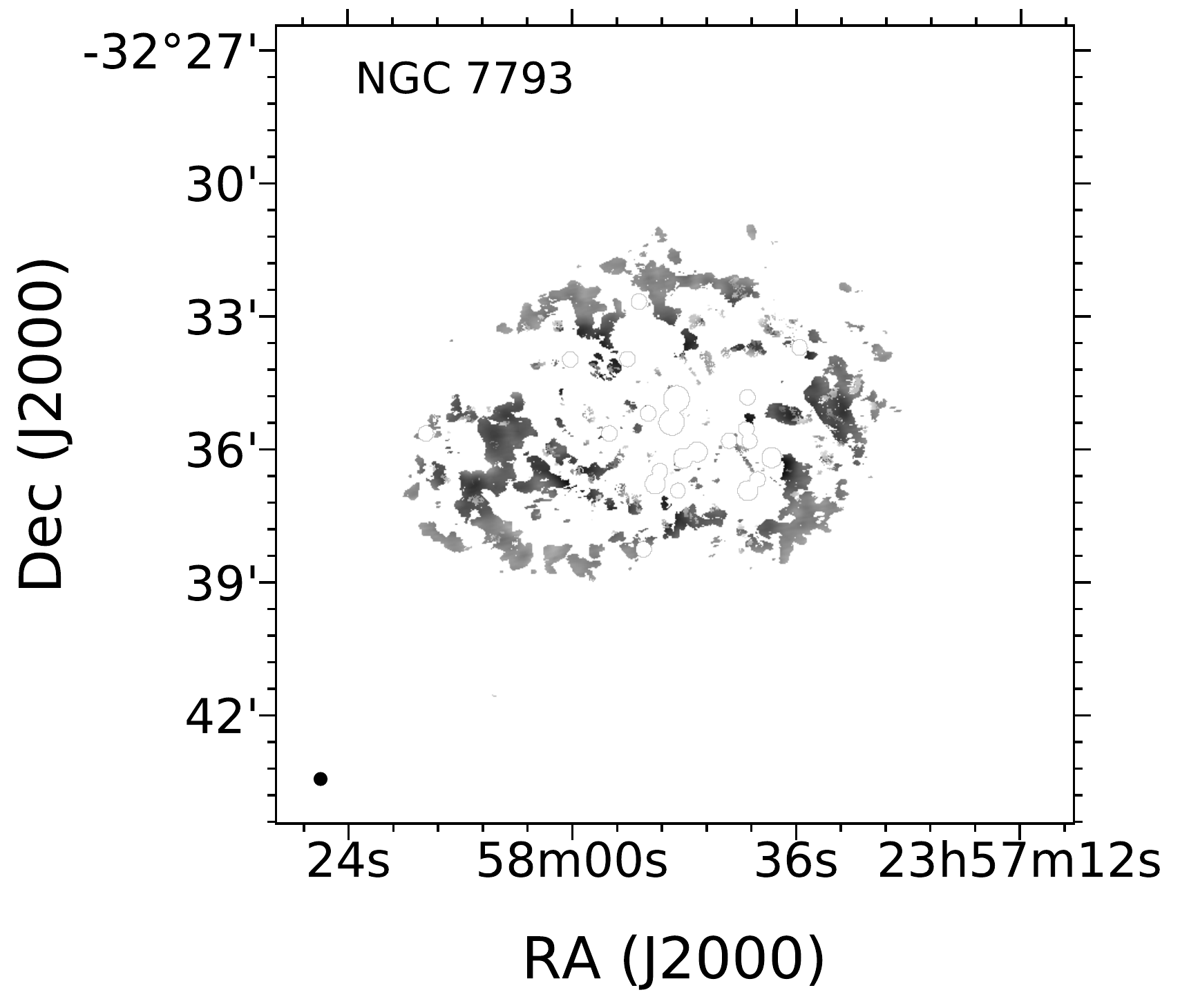}
\includegraphics[width=8cm,height=6cm]{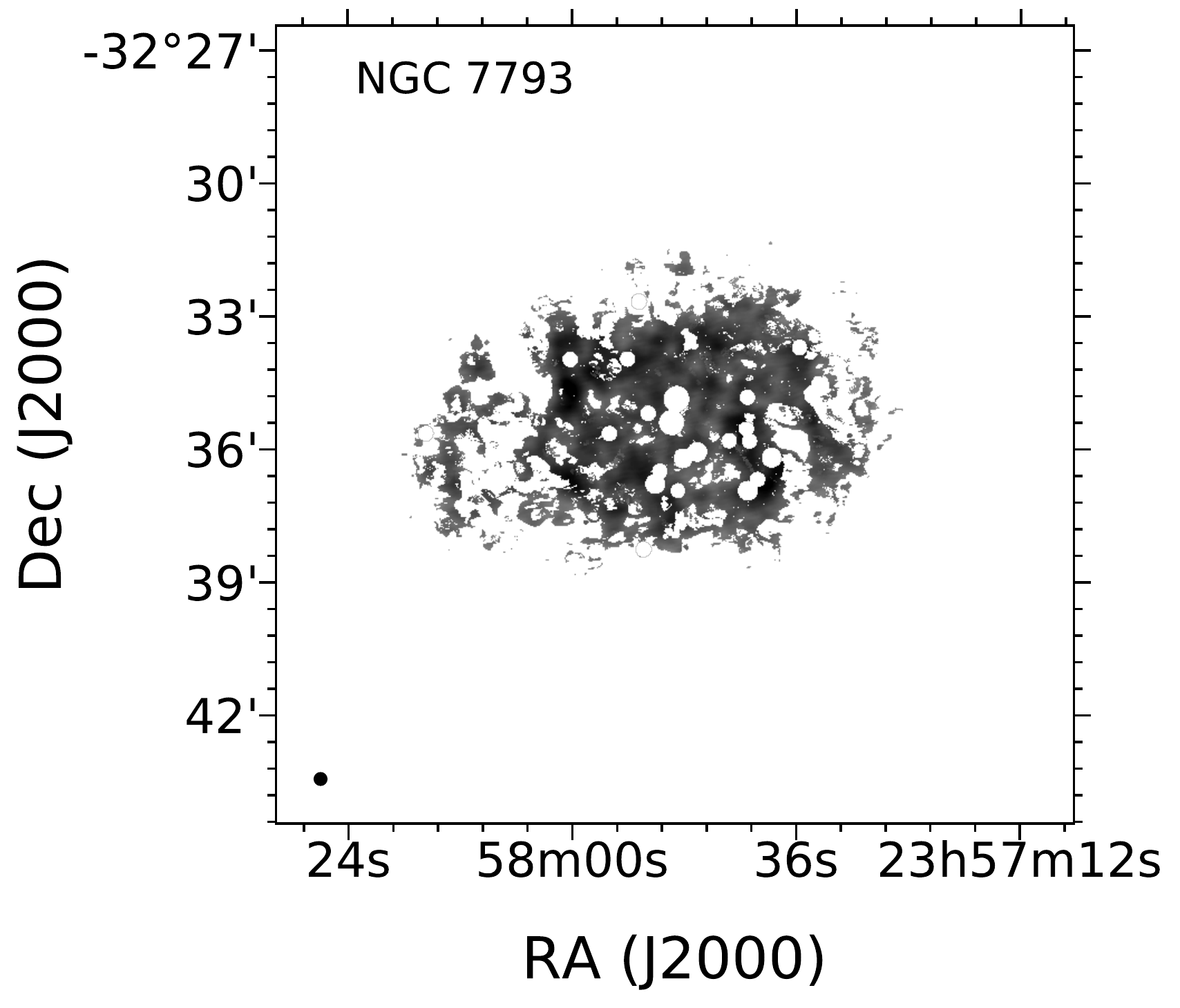}
\caption{The cold (left panel) and warm (right panel) components of HI gas separated into two images using the method described in \citet{Patra+2016}. We see a distinct bi-modality of the atomic gas distribution in NGC 7793. The masks applied to avoid contamination from Galactic foreground stars can be seen as white circles with the convolved beam size (15.6$^{\prime\prime}\times$15.6$^{\prime\prime}$) shown as a black dot in the bottom left corner of each panel. The angular size of these images has been kept the same as Figure \ref{contours_7793} for comparison.}
\label{7HImaps}
\end{figure*}

For NGC 3184, there is no significant blending as the inclination is low (i=16 deg). However, for NGC 7793 (i=50 deg), a significant blending in terms of the long extended tail can be seen. As evident from the histograms, we have used a threshold of $\sigma_{HI}$=10 km/s to separate the wide/narrow components and generate the HI maps individually, for both galaxies. We consider the wide component to trace the ``warm" HI gas and the narrow component to trace the ``cold" HI gas \citep{Young+1997,Young+2003,Begum+2006,Warren+2012}. Hence, in the remainder of this work, we refer to the wide and narrow components as warm and cold HI, respectively. The separated HI maps for the two galaxies are shown in Figures \ref{3HImaps} \& \ref{7HImaps}. \\

\subsection{Data reduction}

\textbf{Moment maps and image data:} The THINGS data are available in the form of data cubes and as individual moment maps. We have used the moment 0 (MOM0) and moment 2 (MOM2) maps which provide integrated HI intensity (Jy/beam*m/s units) and velocity dispersion (m/s) information, respectively. For the IR dust maps, we have used the fifth (and last) data delivery of SINGS for MIR 8 $\mu$m and 24 $\mu$m data. The \textit{Spitzer} image data in \texttt{FITS} format have units of MJy/sr. Since we are using the 8 $\mu$m as a non-stellar PAH emission tracer, we subtract contribution from the stellar continuum using the SINGS 3.6 $\mu$m data following the procedure described in \cite{Helou+2004}. This removes the starlight contribution to the \textit{Spitzer} 8 $\mu$m band but in regions of weak or low PAH emission, there might still be a significant contribution from non-stellar hot dust emission \citep{Bendo+2006}. The 70 $\mu$m, 100 $\mu$m and 160 $\mu$m data have been taken from the second delivery (DR2) of the high-level data products of KINGFISH in units of Jy/pixel.  The CO data from HERACLES have units of K km/s.\\

\textbf{Image convolution and re-sizing:} As seen in Table \ref{table:telescopes}, the radio, IR and CO data available at multiple wavelength bands have different sizes and resolutions. Therefore, in order to bring all the images to equal footing for comparison, we select the largest beam size available for each galaxy in our sample and then convolve the rest of the data to this beam size using standard tasks in the Astronomical Image Processing System (AIPS) software. First, we manually apply masks to all available images such that we can avoid contamination from the Galactic foreground stars, e.g. as seen in Figures \ref{3HImaps} \& \ref{7HImaps}. The images are then convolved such that the output attains the resolution of the largest beam among all the available gas/dust maps. Once all the images for a particular galaxy have been convolved to the new beam size, we re-sample all the image data such that the pixel size becomes equal to the beam size, i.e. one pixel per beam. For example, a radio image with an original beam size of 7.51$^{\prime\prime}\times$6.93$^{\prime\prime}$ for NGC 3184 has 1024$\times$1024 pixels, with a pixel size of 1.5$^{\prime\prime}$ and it has been convolved such that we get a final output beam size of 13.4$^{\prime\prime}\times$13.4$^{\prime\prime}$, which is in accordance with the highest beam size in the sample belonging to the CO map. We divide the beam size by the pixel size (13.4/1.5) to get the number of pixels in a beam (=8.9) and then divide the total number of pixels in the image by this number (1024/8.9) to get the new number of image pixels (=115$\times$115), for re-sampling to make the pixel size and beam size equal. Moreover, the size of the images and their reference pixels need to be re-aligned since our objective is to do a pixel-by-pixel comparative study and this has been done using the Common Astronomy Software Applications (CASA) package. Thus, we bring all the images (including the cold/warm HI maps) to a uniform size and geometry for meaningful comparison.\\

\begin{figure*}
\centering
\includegraphics[width=8cm,height=6cm]{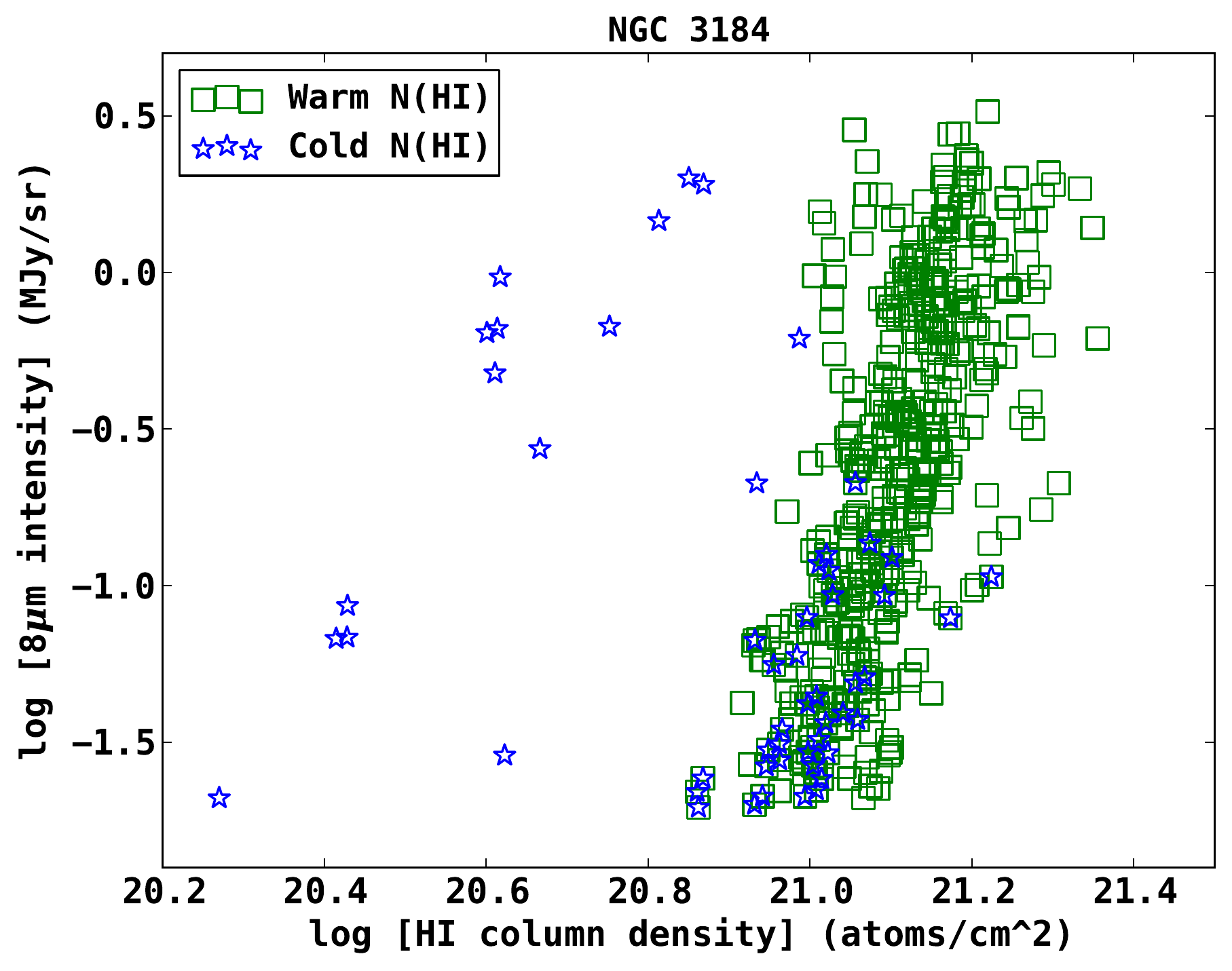}
\includegraphics[width=8cm,height=6cm]{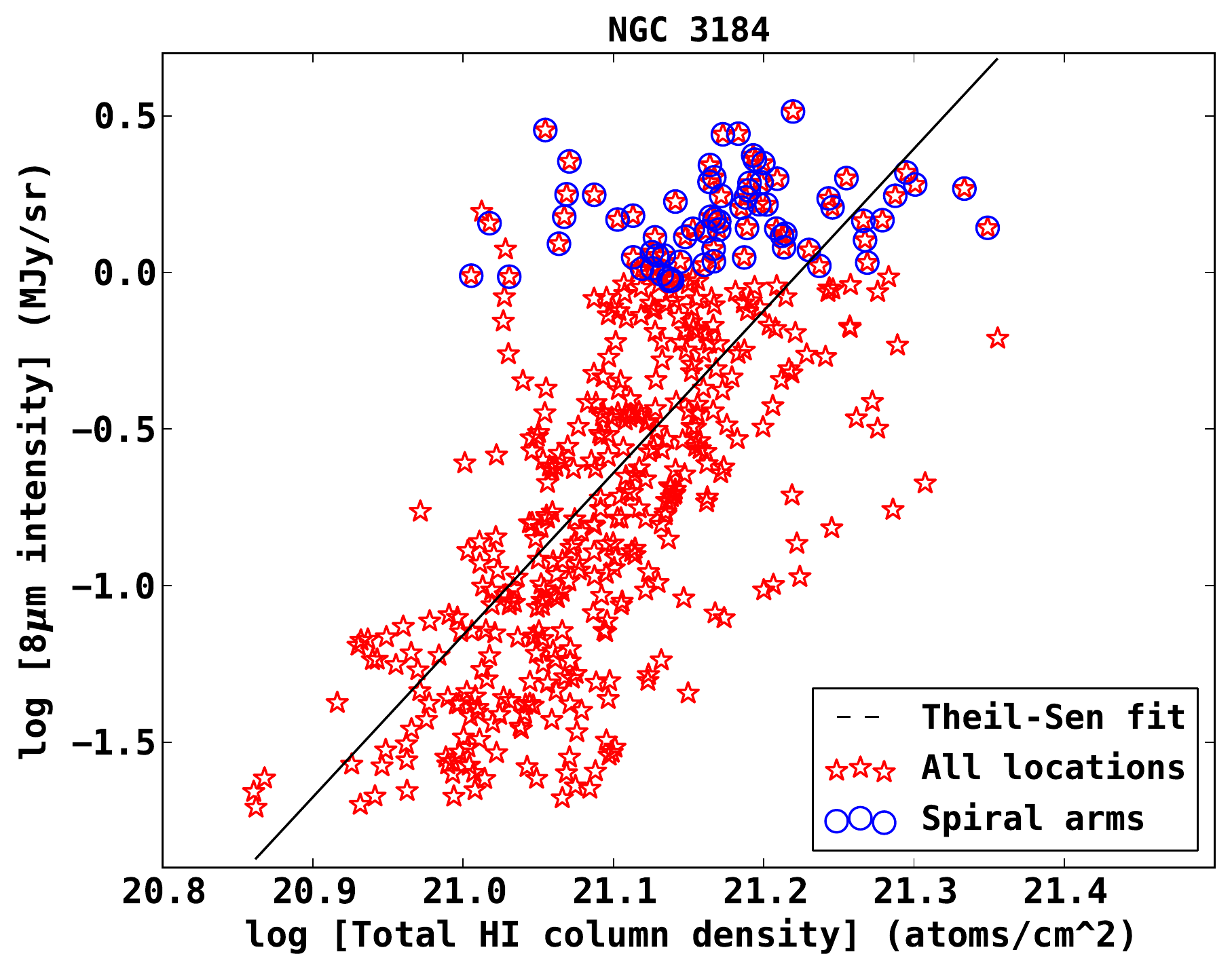}
\includegraphics[width=8cm,height=6cm]{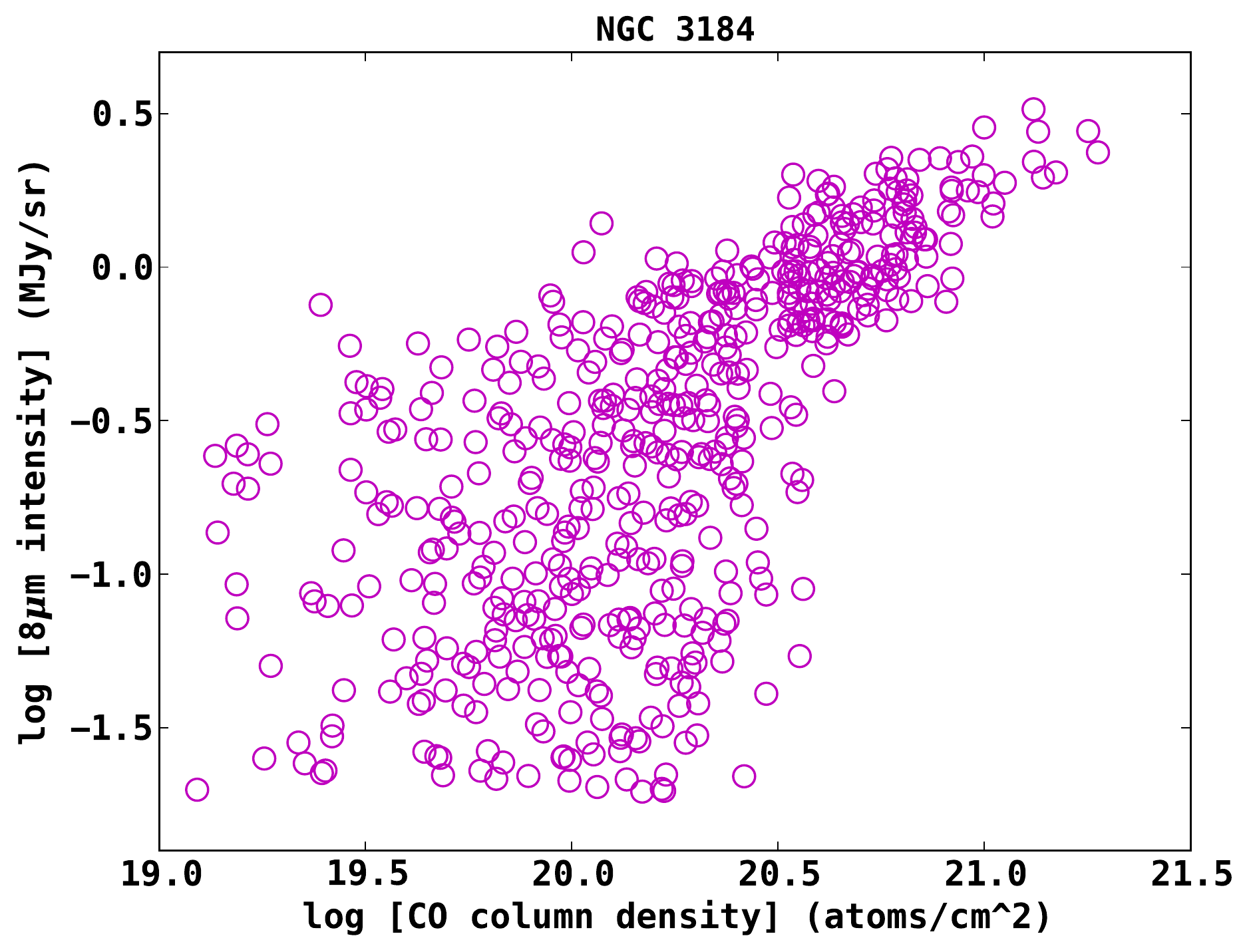}
\includegraphics[width=8cm,height=6cm]{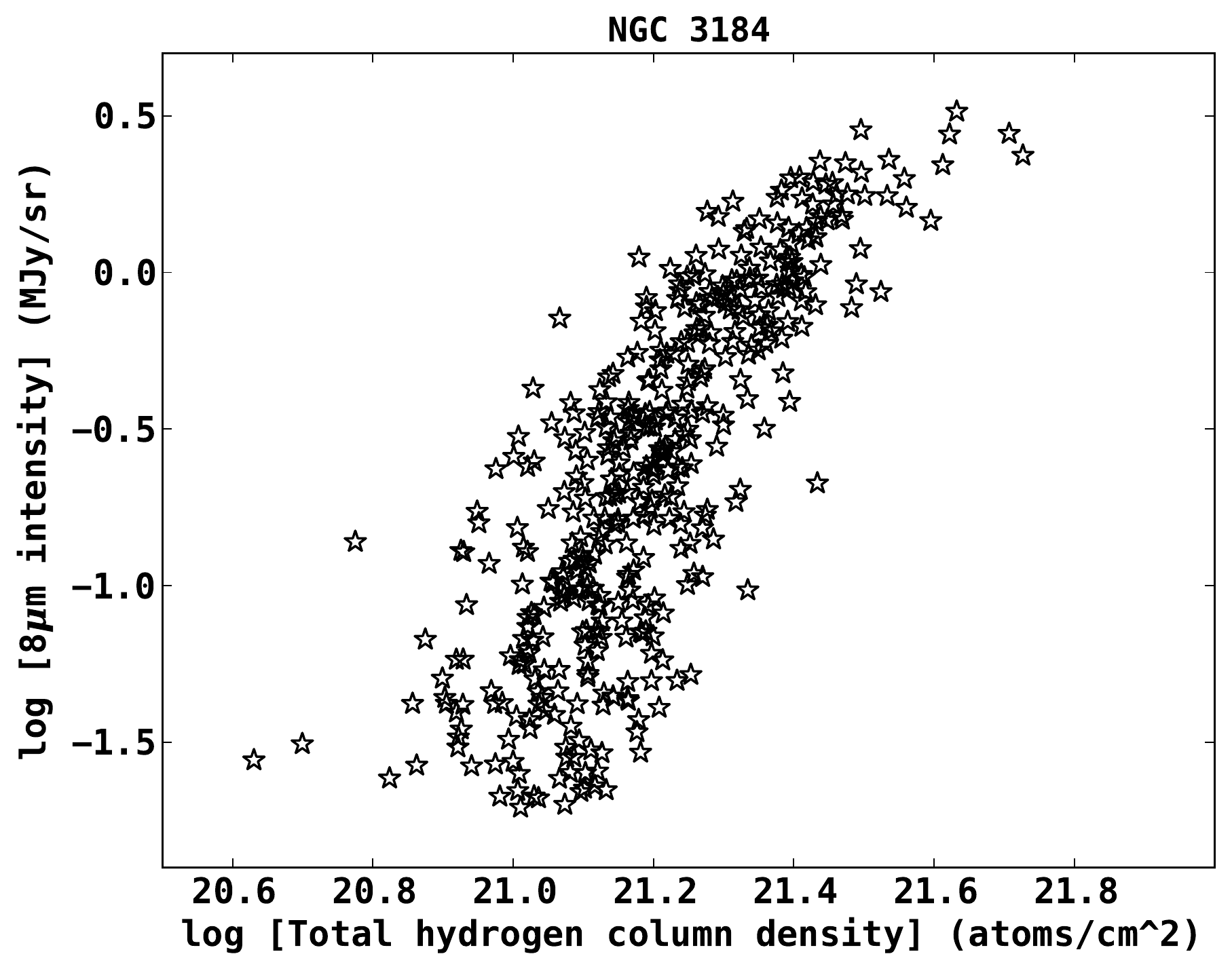}
\caption{Gas vs dust (8 $\mu$m) correlation plots for \textbf{NGC 3184}. The top left panel shows the cold/warm gas vs dust, top right panel shows total atomic HI gas vs dust, the bottom left panel shows molecular gas (CO) vs dust and the bottom right panel shows total (atomic+molecular) gas vs dust correlation. The tightest correlation is seen in the total gas vs dust case. The locations of spiral arm emission have also been marked on top of the total atomic HI gas vs dust correlation plot (as blue circles). The dashed line (in black) in the same plot shows the Theil-Sen regression fit which is insensitive to outliers.}
\label{3plots_irHI8}
\end{figure*}

\begin{figure*}
\centering
\includegraphics[width=8cm,height=6cm]{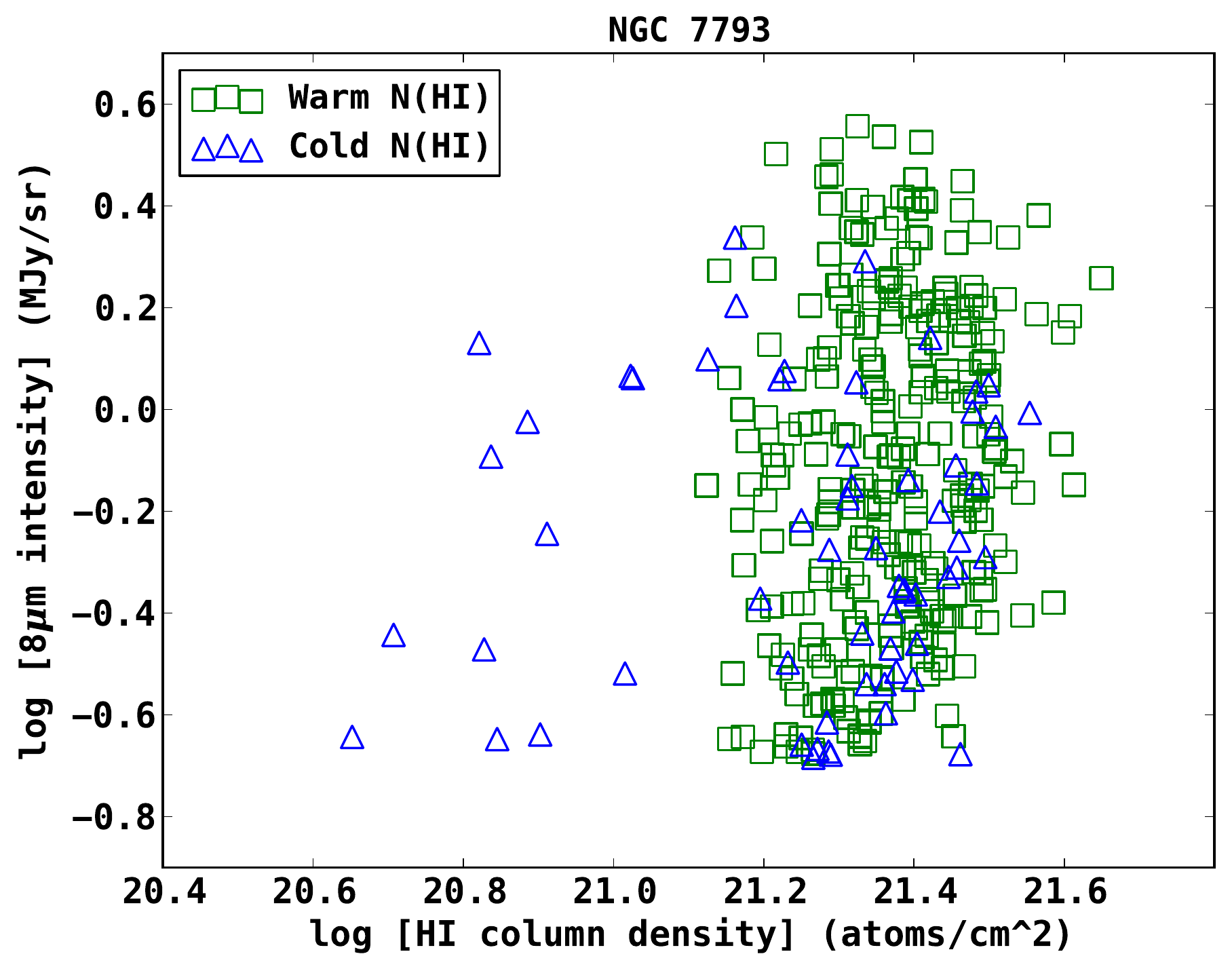}
\includegraphics[width=8cm,height=6cm]{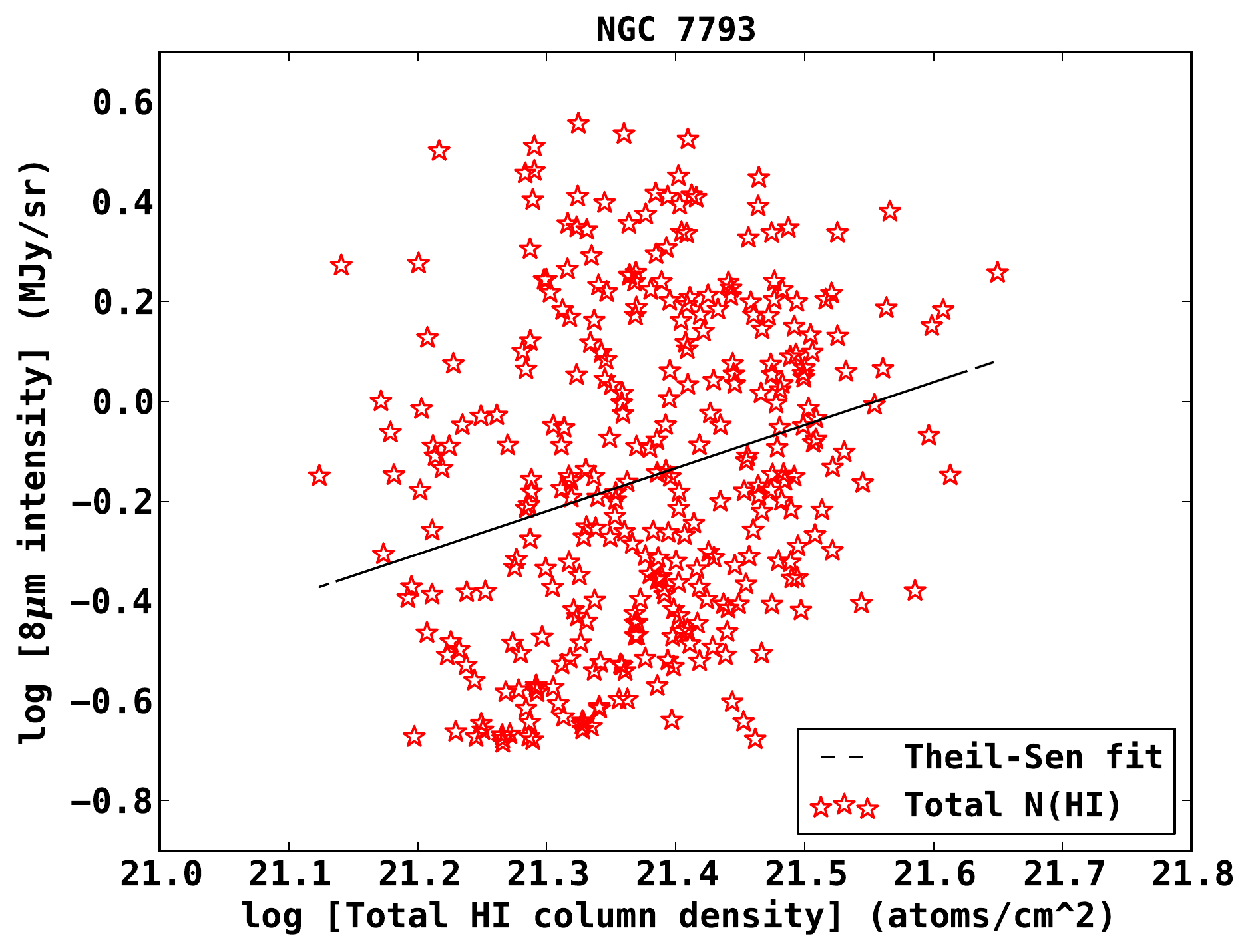}
\caption{Gas vs dust (8 $\mu$m) correlation plots for \textbf{NGC 7793}. The cold/warm gas vs dust correlation is shown in the left panel while the total atomic HI gas vs dust is shown in the right panel. We have shown the plots for the MIR 8 $\mu$m case which is weakly correlated to the gas. It is to be noted that this correlation gets tighter as we move towards the FIR in NGC 7793. The dashed line (in black) in the right panel shows the Theil-Sen regression fit which is insensitive to outliers.}
\label{7plots_irHI8}
\end{figure*}

\begin{figure*}
\centering
\includegraphics[width=8cm,height=6cm]{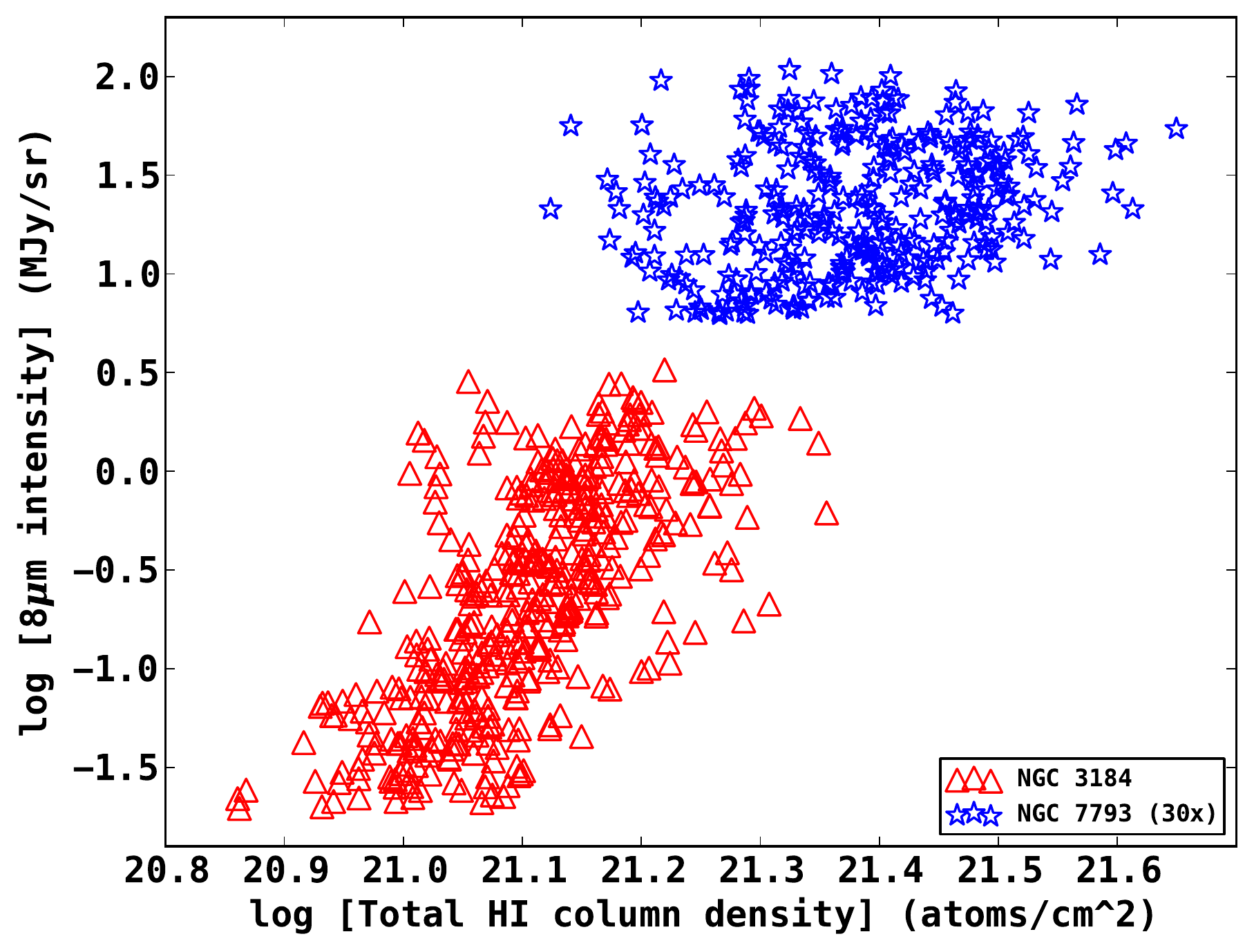}
\includegraphics[width=8cm,height=6cm]{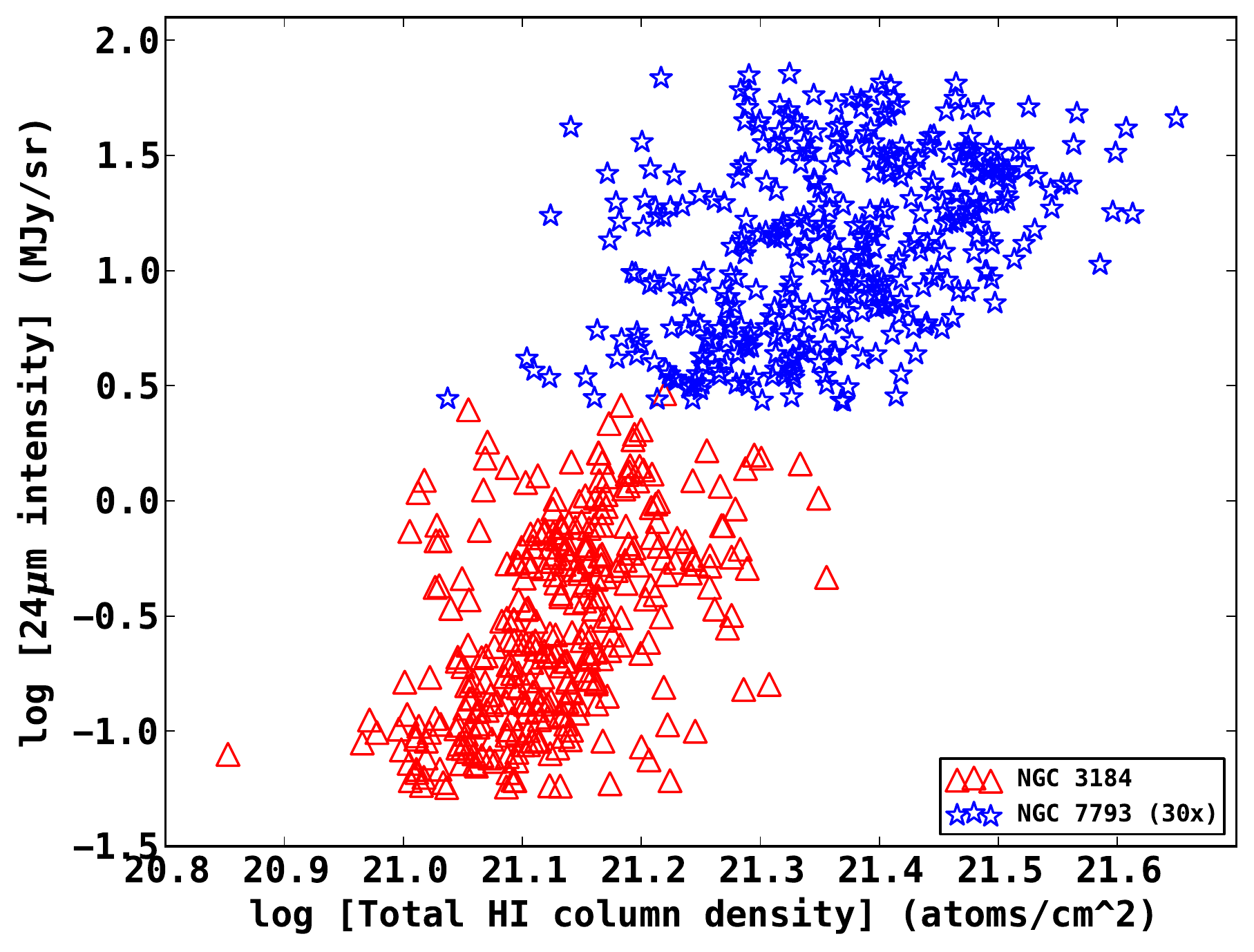}
\includegraphics[width=8cm,height=6cm]{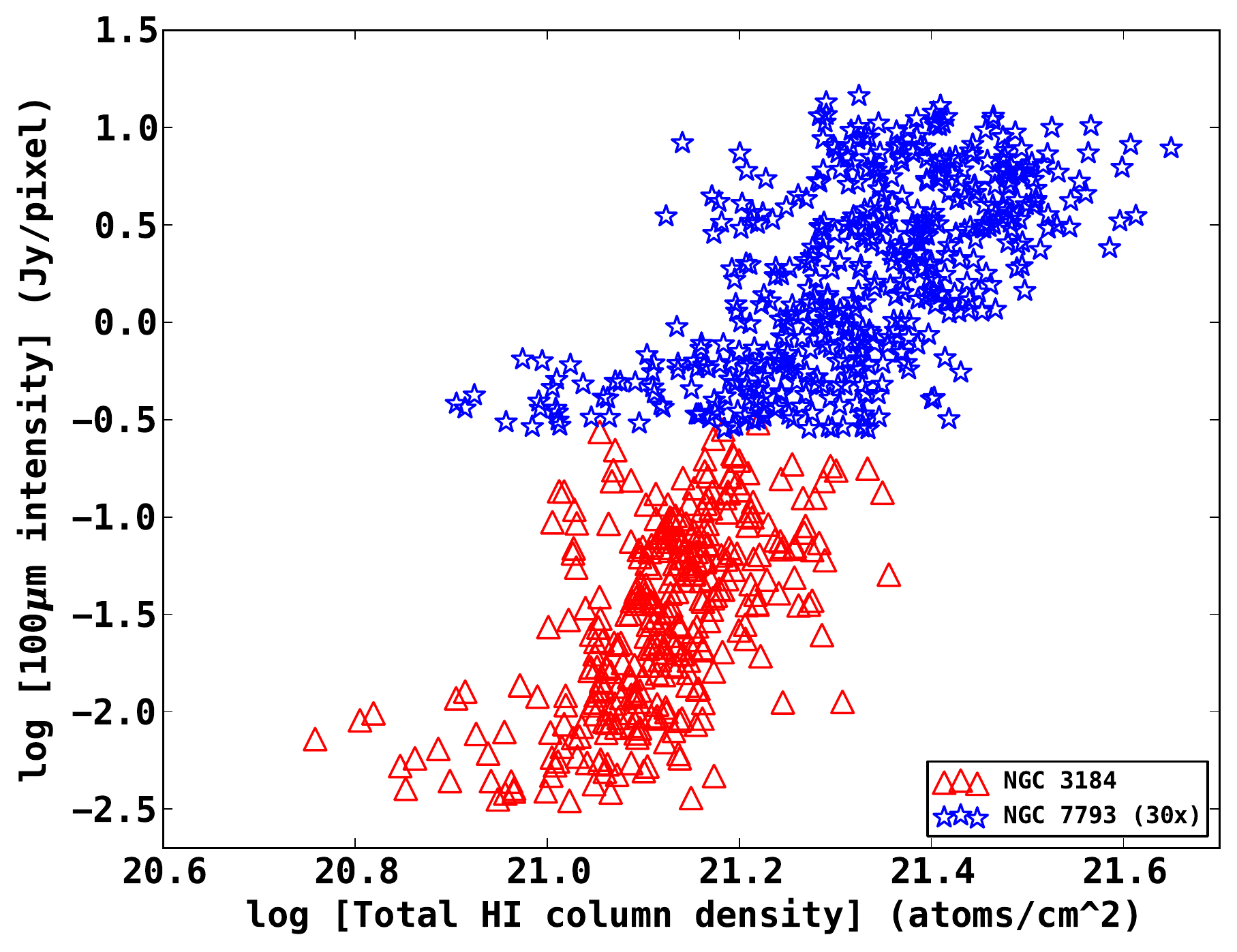}
\includegraphics[width=8cm,height=6cm]{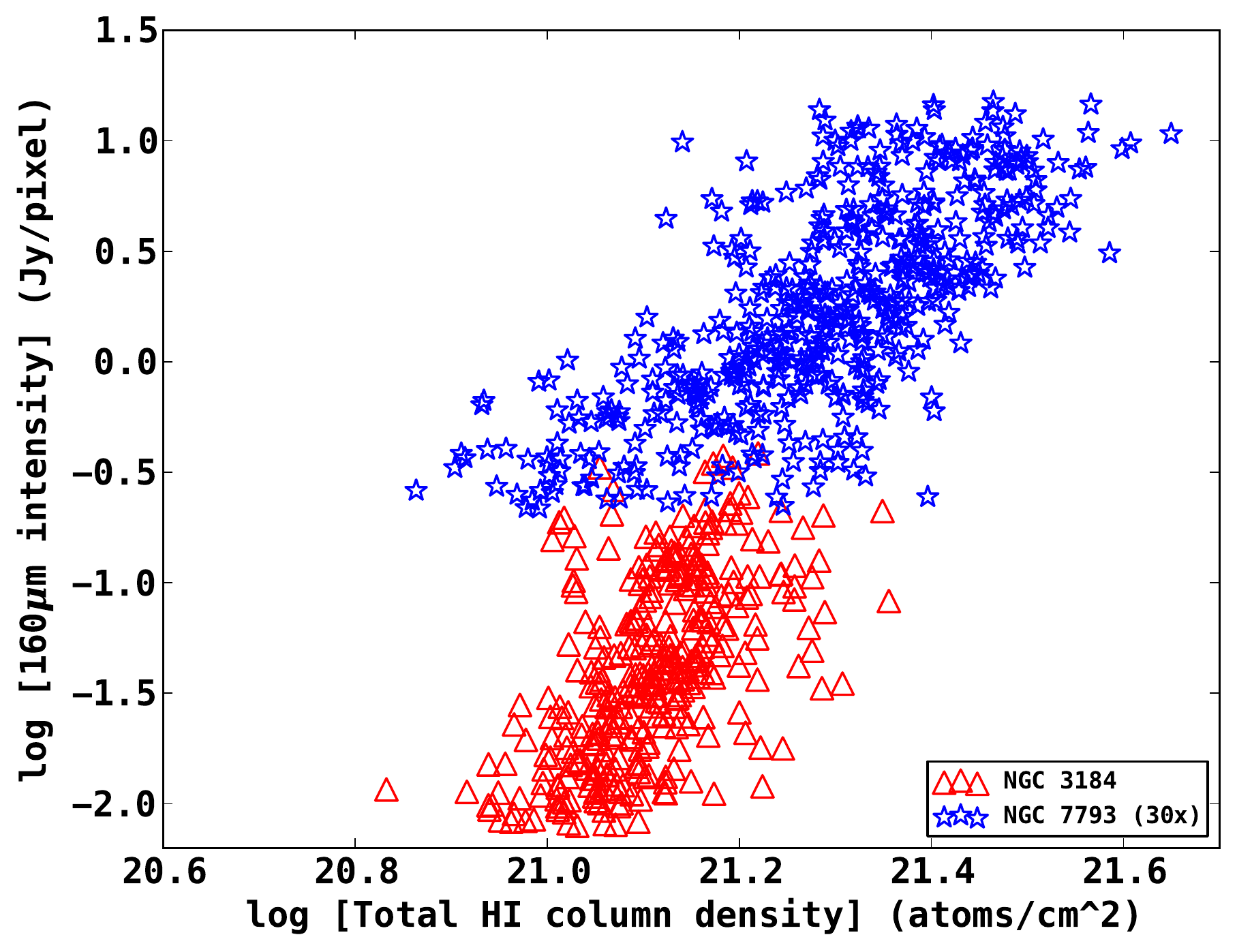}
\caption{Overlaid plots for N(HI) gas vs dust IR at MIR 8 $\mu$m (top left panel), MIR 24 $\mu$m (top right panel), FIR 100 $\mu$m (bottom left panel) and FIR 160 $\mu$m (bottom right panel). The lack of MIR emission in NGC 7793 is clearly visible from the top two panels. There is also a distinct difference in the atomic gas-to-dust ratio between the two. Here, in each panel, we have multiplied a factor of $\sim$30 times the original IR emission in NGC 7793 to scale it up to the same correlation as NGC 3184. This shows a much higher fraction of atomic gas or lesser fraction of dust in NGC 7793 as compared to NGC 3184.}
\label{plots_overlay}
\end{figure*}

\section{Results \& Discussion}
\label{results}

We have carried out a pixel-by-pixel correlation study among the different available radio, CO and IR bands for both the galaxies. We consider a noise (rms) in each IR image, away from the optical disk of the galaxy region such that it does not overlap with any bright source and is purely from the foreground/background contribution, and then set a threshold of 2.5*rms as the signal-to-noise cut-off. For the HI observations, we have converted the units of the image into atomic column density units [N(HI)] by setting a minimum cut-off at 5$\times$10$^{19}$ atoms/cm$^{2}$ according to the sensitivity considerations of THINGS survey \citep{Walter+2008}. In case of the CO data \citep{Leroy+2009}, we have used a line ratio of R$_{21}$=(2$\rightarrow$1)/(1$\rightarrow$0)=0.7 \citep{Sandstrom+2013} and an X$_{CO}$=2$\times$10$^{20}$ cm$^{-2}$(K km/s)$^{-1}$ factor \citep{Bolatto+2013} to obtain the molecular hydrogen column densities [N(H$_2$)]. \\

Figures \ref{3plots_irHI8} \& \ref{7plots_irHI8} show the gas-dust correlations for the 8 $\mu$m dust band in NGC 3184 and NGC 7793, respectively. The four panels in Figure \ref{3plots_irHI8} correspond to the separated warm/cold HI gas, total HI gas, molecular CO gas and total (atomic+molecular) gas column densities with respect to 8 $\mu$m dust emission in NGC 3184. The spiral arm locations have been marked separately in the total gas-dust correlation plot for NGC 3184 which have been identified using a combination of 3.6 $\mu$m and H$\alpha$ images available in the SINGS data archive \citep{Kennicutt+2003}. Since the CO data was unavailable for NGC 7793 in the public archives, we have only shown the separated warm/cold HI and the total HI gas column density vs 8 $\mu$m dust emission for this galaxy in Figure \ref{7plots_irHI8}. Being a flocculent spiral, the arms in NGC 7793 are distributed throughout the galaxy and hence we haven't been able to separate them from the inter-arm locations. While we have not shown all the gas-dust correlation plots for individual IR bands here, we have used the non-parametric Spearman's ($\rho$) and Kendall's ($\tau$) rank correlation coefficients to quantify the relationship between various observed parameters and the coefficient values are presented in Tables \ref{corr_irHI} \& \ref{corr_mfir}. In addition, we have compared the total N(HI) gas-dust correlations for the two galaxies at different IR bands as shown in Figure \ref{plots_overlay}. \\

The results show that the trend in the variation of gas-dust correlations with individual IR bands is different for each galaxy. While the N(HI) and dust IR bands are positively correlated with almost similar values at each individual band for NGC 3184, the values of correlation coefficients increase drastically with an increase in IR emission wavelength for NGC 7793. In NGC 3184, the Spearman's rank correlation coefficient changes from $\sim$0.7 at N(HI) vs MIR (8 $\mu$m) to $\sim$0.6 at N(HI) vs FIR (160 $\mu$m). On the other hand, the same coefficient value changes from $\sim$0.2 at N(HI) vs MIR (8 $\mu$m) to $\sim$0.7 at N(HI) vs FIR (160 $\mu$m) in NGC 7793. The trend remains the same even when we separate the atomic HI into warm and cold components to check the correlations individually. A similar situation was observed in the case of M33 by \cite{Hinz+2004}, who found the 24 and 70 $\mu$m emissions to trace highly ionizing stars but the 160 $\mu$m emission to be more closely related to cool, diffuse dust which matches the non-thermal radio emission distribution. With the similarity in properties between M33 and NGC 7793, this indicates that the dust population responsible for gas-dust interactions in NGC 7793 has a higher contribution from larger, relatively cooler dust components that show emission towards the FIR. This could also indicate a deficiency in smaller-sized dust particles in the flocculent spiral NGC 7793 due to their possible destruction in the vicinity of highly ionizing sources.\\

We see that the cold HI is relatively weakly correlated with the dust emission as compared to the warm HI gas (Table \ref{corr_irHI}). This means that the total N(HI) correlation which is seen with individual dust bands is mainly due to the warm gas component in both galaxies (Figures \ref{3plots_irHI8} \& \ref{7plots_irHI8}). The molecular gas in NGC 3184 shows a good positive correlation ($\sim$0.7) with all IR bands and it also dominates the total gas (atomic+molecular) vs dust correlation which is tighter ($\sim$0.8) than either individual case as seen in Figure \ref{3plots_irHI8}. If we only look at the spiral arm locations in NGC 3184, we do not see any correlation due to the narrow range of IR emission values covered. We see the overall positive correlation only when we consider all locations. Moreover, the spiral arms do not have any preferential cold/warm gas dominance in NGC 3184. These locations are, of course, concentrated towards higher dust emission regions due to the higher dust concentration in the spiral arms (Figure \ref{3plots_irHI8}).\\

When we look at the MIR vs FIR correlations in both galaxies, we find an overall similar positive correlation trend between the different dust bands indicating a mixture of hot and cold dust components. While the correlation coefficients among the different 8, 24, 100 \& 160 $\mu$m emissions have similar values for both galaxies, there is a slight difference in case of the correlation coefficients involving 70 $\mu$m emission. For instance, the 70 vs 100 $\mu$m Spearman's rank correlation coefficient for NGC 3184 is $\sim$0.6 while it is $\sim$0.8 for NGC 7793 (Table \ref{corr_mfir}). So, while a large fraction of the dust particles responsible for the 70 $\mu$m emission may come from similar grains which show emission at 24 $\mu$m with sensitivity to star-formation effects \citep{Dale+2005,Calzetti+2005}, the 24 $\mu$m may also arise from diffuse cirrus especially in inter-arm regions \citep{Foyle+2010}. Therefore, the relatively weaker correlation between other dust bands and 70 $\mu$m  emission in NGC 3184 as compared to NGC 7793 might be due to higher contribution towards the cold dust emission from the inter-arm regions at the other IR bands as opposed to the warm dust emission at 70 $\mu$m. These observations imply that NGC 7793 has a higher cold dust fraction as compared to NGC 3184.\\

In Figure \ref{plots_overlay}, we have shown the correlation between the total neutral atomic gas and the dust IR emission for NGC 3184 and NGC 7793 in the same panels for comparison. We find the 8 $\mu$m dust emission in NGC 7793 to be more concentrated towards the region showing spiral arms in NGC 3184 which is because of the uniformly distributed nature of spiral arms in the flocculent spiral NGC 7793, which are difficult to visually distinguish. The lack of dust emission at 8 $\mu$m could also indicate the dissociation/destruction of smaller-sized dust particles in the inter-arm regions of NGC 7793. The existence of more widespread emission towards 160 $\mu$m in NGC 7793 supports this as larger-sized dust particles are not easily destroyed in similar harsh environments. Moreover, we see that in the case of NGC 7793, there is a greater amount of neutral gas N(HI) for the same amount of dust which is present in NGC 3184. The dust-to-neutral atomic gas N(HI) ratio is found to be roughly 30 times lower than that in NGC 3184 as seen in Figure \ref{plots_overlay}. \cite{Draine+2007SINGS} investigated the dust-to-gas mass ratio (DGR) within the SINGS galaxy sample and found the DGR variation to be much more dependent on a change in metallicity as compared to Hubble type, which seems to imply that the variation seen in our case is also due to metallicity effects rather than galaxy type. In addition, we also found the FIR/MIR ratio to be higher in NGC 7793 by a factor of $\sim$1.6 times that in NGC 3184, further supporting the existence of a higher fraction of larger-sized particles which show FIR emission in NGC 7793.

\section{Conclusions}
\label{conclusions}

We report the results of a pixel-by-pixel gas-dust correlation study in two spiral galaxies: NGC 3184 and NGC 7793, based on an analysis of archival data. For the first time, we have separated the narrow (cold) and wide (warm) components of atomic HI gas in these two galaxies and investigated their correlations separately with individual dust IR bands, in a spatially resolved way. We have also compared the findings for these two galaxies and the main conclusions are summarized below:
\begin{itemize}

\item We find that the HI gas distribution in both NGC 3184 and NGC 7793 clearly shows a bi-modality with distinct differences in the cold and warm gas components, which we have identified and separated.\\

\item The atomic gas shows an overall positive correlation with the dust infrared emission bands in both the galaxies. On decomposing the atomic gas into cold/warm components, we find that the total N(HI) correlation does not have much contribution from the cold gas in either case. \\

\item The spiral arm locations in NGC 3184 do not show any preferential dominance of the warm or cold gas components. Moreover, the dust vs total (atomic+molecular) gas shows a very good correlation in NGC 3184, with the total gas-dust interactions being heavily influenced by the molecular gas phase. The unavailability of required data for NGC 7793 limits us from obtaining its molecular gas to dust correlations. \\

\item We also seem to find a distinct difference in the particle size distribution of dust within these two galaxies, as summarized below: \\

\item[$-$] The dust to atomic gas N(HI) ratio is found to be much lower in NGC 7793 with respect to NGC 3184, which indicates a higher concentration of atomic gas or lower dust emission in NGC 7793. A lack of dust emission in the MIR, coupled with the higher FIR/MIR ratio found in NGC 7793 as compared to NGC 3184 indicates a deficiency in smaller-sized grains with most of the observed FIR emission in NGC 7793 being contributed by larger-sized dust particles.  \\

\item[$-$] The gas-dust correlation trend at different MIR and FIR bands in NGC 3184 is quite uniform while that in NGC 7793 shows a much better gas-dust correlation towards the FIR. This indicates that the dust in NGC 3184 is a mixture of both hot and cold grains, while that in the flocculent spiral galaxy NGC 7793 has a dominance of cold and diffuse, larger-sized dust over smaller-sized dust particles.\\

\end{itemize}

From our ongoing and future work on the full sample of galaxies, we hope to understand the physical significance of dust much better than what has been accomplished in this pilot study. We expect that a larger galaxy sample will make the differences in gas-dust correlations among different galaxy types more clear.

\section*{Acknowledgements}

We thank the anonymous reviewer for useful comments and suggestions towards improvement of this study. This work made use of THINGS, `The HI Nearby Galaxy Survey' \citep{Walter+2008} and HERACLES, `The HERA CO-Line Extragalactic Survey' \citep{Leroy+2009}. This research has made use of the NASA/IPAC Infrared Science Archive and the NASA/IPAC Extragalactic Database (NED), which are operated by the Jet Propulsion Laboratory, California Institute of Technology, under contract with the National Aeronautics and Space Administration. This research has made use of the SIMBAD database, operated at CDS, Strasbourg, France.\\

NR acknowledges support from the Infosys Foundation through the Infosys Young Investigator grant, and from the Max-Plack-Gesellschaft through the Max Planck Partner Group grant. CJ would like to thank the DST, Government of India for support via a J.C. Bose fellowship (SB/S2/JCB31/2014).




\bibliographystyle{mnras}
\bibliography{mybib} 

\begin{thebibliography}{}
\makeatletter
\relax
\def\mn@urlcharsother{\let\do\@makeother \do\$\do\&\do\#\do\^\do\_\do\%\do\~}
\def\mn@doi{\begingroup\mn@urlcharsother \@ifnextchar [ {\mn@doi@}
  {\mn@doi@[]}}
\def\mn@doi@[#1]#2{\def\@tempa{#1}\ifx\@tempa\@empty \href
  {http://dx.doi.org/#2} {doi:#2}\else \href {http://dx.doi.org/#2} {#1}\fi
  \endgroup}
\def\mn@eprint#1#2{\mn@eprint@#1:#2::\@nil}
\def\mn@eprint@arXiv#1{\href {http://arxiv.org/abs/#1} {{\tt arXiv:#1}}}
\def\mn@eprint@dblp#1{\href {http://dblp.uni-trier.de/rec/bibtex/#1.xml}
  {dblp:#1}}
\def\mn@eprint@#1:#2:#3:#4\@nil{\def\@tempa {#1}\def\@tempb {#2}\def\@tempc
  {#3}\ifx \@tempc \@empty \let \@tempc \@tempb \let \@tempb \@tempa \fi \ifx
  \@tempb \@empty \def\@tempb {arXiv}\fi \@ifundefined
  {mn@eprint@\@tempb}{\@tempb:\@tempc}{\expandafter \expandafter \csname
  mn@eprint@\@tempb\endcsname \expandafter{\@tempc}}}

\bibitem[\protect\citeauthoryear{{Abdullah} et~al.,}{{Abdullah}
  et~al.}{2017}]{Abdullah+2017}
{Abdullah} A.,  et~al., 2017, \mn@doi [\apj] {10.3847/1538-4357/aa6fa9}, \href
  {http://adsabs.harvard.edu/abs/2017ApJ...842....4A} {842, 4}

\bibitem[\protect\citeauthoryear{{Begum}, {Chengalur}  \& {Bhardwaj}}{{Begum}
  et~al.}{2006}]{Begum+2006}
{Begum} A.,  {Chengalur} J.~N.,   {Bhardwaj} S.,  2006, \mn@doi [\mnras]
  {10.1111/j.1745-3933.2006.00220.x}, \href
  {https://ui.adsabs.harvard.edu/abs/2006MNRAS.372L..33B} {372, L33}

\bibitem[\protect\citeauthoryear{{Bendo} et~al.,}{{Bendo}
  et~al.}{2006}]{Bendo+2006}
{Bendo} G.~J.,  et~al., 2006, \mn@doi [\apj] {10.1086/508057}, \href
  {http://adsabs.harvard.edu/abs/2006ApJ...652..283B} {652, 283}

\bibitem[\protect\citeauthoryear{{Bigiel}, {Leroy}, {Walter}, {Brinks}, {de
  Blok}, {Madore}  \& {Thornley}}{{Bigiel} et~al.}{2008}]{Bigiel+2008}
{Bigiel} F.,  {Leroy} A.,  {Walter} F.,  {Brinks} E.,  {de Blok} W.~J.~G.,
  {Madore} B.,   {Thornley} M.~D.,  2008, \mn@doi [\aj]
  {10.1088/0004-6256/136/6/2846}, \href
  {http://adsabs.harvard.edu/abs/2008AJ....136.2846B} {136, 2846}

\bibitem[\protect\citeauthoryear{{Bolatto}, {Wolfire}  \& {Leroy}}{{Bolatto}
  et~al.}{2013}]{Bolatto+2013}
{Bolatto} A.~D.,  {Wolfire} M.,   {Leroy} A.~K.,  2013, \mn@doi [Annual Review
  of Astronomy and Astrophysics] {10.1146/annurev-astro-082812-140944}, \href
  {https://ui.adsabs.harvard.edu/abs/2013ARA&A..51..207B} {51, 207}

\bibitem[\protect\citeauthoryear{Braine, Combes, Casoli, Dupraz, G{\'e}rin,
  Klein, Wielebinski  \& Brouillet}{Braine et~al.}{1993}]{Braine+1993}
Braine J.,  Combes F.,  Casoli F.,  Dupraz C.,  G{\'e}rin M.,  Klein U.,
  Wielebinski R.,   Brouillet N.,  1993, Astronomy and Astrophysics Supplement
  Series, 97, 887

\bibitem[\protect\citeauthoryear{{Calzetti} et~al.,}{{Calzetti}
  et~al.}{2005}]{Calzetti+2005}
{Calzetti} D.,  et~al., 2005, \mn@doi [The Astrophysical Journal]
  {10.1086/466518}, \href {http://adsabs.harvard.edu/abs/2005ApJ...633..871C}
  {633, 871}

\bibitem[\protect\citeauthoryear{{Calzetti} et~al.,}{{Calzetti}
  et~al.}{2007}]{Calzetti+2007}
{Calzetti} D.,  et~al., 2007, \mn@doi [The Astrophysical Journal]
  {10.1086/520082}, \href {http://adsabs.harvard.edu/abs/2007ApJ...666..870C}
  {666, 870}

\bibitem[\protect\citeauthoryear{{Dale} et~al.,}{{Dale}
  et~al.}{2005}]{Dale+2005}
{Dale} D.~A.,  et~al., 2005, \mn@doi [The Astrophysical Journal]
  {10.1086/491642}, \href {http://adsabs.harvard.edu/abs/2005ApJ...633..857D}
  {633, 857}

\bibitem[\protect\citeauthoryear{{Doane}, {Sanders}, {Wilcots}  \&
  {Juda}}{{Doane} et~al.}{2004}]{Doane+2004}
{Doane} N.~E.,  {Sanders} W.~T.,  {Wilcots} E.~M.,   {Juda} M.,  2004, \mn@doi
  [\aj] {10.1086/425627}, \href
  {http://adsabs.harvard.edu/abs/2004AJ....128.2712D} {128, 2712}

\bibitem[\protect\citeauthoryear{{Draine}}{{Draine}}{2003}]{Draine+2003}
{Draine} B.~T.,  2003, \mn@doi [Annual Review of Astronomy and Astrophysics]
  {10.1146/annurev.astro.41.011802.094840}, \href
  {http://adsabs.harvard.edu/abs/2003ARA%26A..41..241D} {41, 241}

\bibitem[\protect\citeauthoryear{{Draine}}{{Draine}}{2011}]{Draine+2011}
{Draine} B.~T.,  2011, {Physics of the Interstellar and Intergalactic Medium}

\bibitem[\protect\citeauthoryear{{Draine} et~al.,}{{Draine}
  et~al.}{2007}]{Draine+2007SINGS}
{Draine} B.~T.,  et~al., 2007, \mn@doi [The Astrophysical Journal]
  {10.1086/518306}, \href
  {https://ui.adsabs.harvard.edu/abs/2007ApJ...663..866D} {663, 866}

\bibitem[\protect\citeauthoryear{{Dyson} \& {Williams}}{{Dyson} \&
  {Williams}}{1997}]{Dyson+1997}
{Dyson} J.~E.,  {Williams} D.~A.,  1997, {The physics of the interstellar
  medium}, \mn@doi{10.1201/9780585368115.
}

\bibitem[\protect\citeauthoryear{{Elmegreen} \& {Elmegreen}}{{Elmegreen} \&
  {Elmegreen}}{1982}]{Elmegreen+1982}
{Elmegreen} D.~M.,  {Elmegreen} B.~G.,  1982, \mn@doi [Monthly Notices of the
  Royal Astronomical Society] {10.1093/mnras/201.4.1021}, \href
  {https://ui.adsabs.harvard.edu/abs/1982MNRAS.201.1021E} {201, 1021}

\bibitem[\protect\citeauthoryear{{Fazio} et~al.,}{{Fazio}
  et~al.}{2004}]{Fazio+2004}
{Fazio} G.~G.,  et~al., 2004, \mn@doi [The Astrophysical Journals]
  {10.1086/422843}, \href {http://adsabs.harvard.edu/abs/2004ApJS..154...10F}
  {154, 10}

\bibitem[\protect\citeauthoryear{Flower \& Pineau Des~For{\^e}ts}{Flower \&
  Pineau Des~For{\^e}ts}{2010}]{Flower+2010}
Flower D.,  Pineau Des~For{\^e}ts G.,  2010, Monthly Notices of the Royal
  Astronomical Society, 406, 1745

\bibitem[\protect\citeauthoryear{{Foyle}, {Rix}, {Walter}  \& {Leroy}}{{Foyle}
  et~al.}{2010}]{Foyle+2010}
{Foyle} K.,  {Rix} H.-W.,  {Walter} F.,   {Leroy} A.~K.,  2010, \mn@doi [\apj]
  {10.1088/0004-637X/725/1/534}, \href
  {http://adsabs.harvard.edu/abs/2010ApJ...725..534F} {725, 534}

\bibitem[\protect\citeauthoryear{{Gil de Paz} et~al.,}{{Gil de Paz}
  et~al.}{2007}]{GildePaz+2007}
{Gil de Paz} A.,  et~al., 2007, \mn@doi [\apjs] {10.1086/516636}, \href
  {http://adsabs.harvard.edu/abs/2007ApJS..173..185G} {173, 185}

\bibitem[\protect\citeauthoryear{{Helfer}, {Thornley}, {Regan}, {Wong},
  {Sheth}, {Vogel}, {Blitz}  \& {Bock}}{{Helfer} et~al.}{2003}]{Helfer+2003}
{Helfer} T.~T.,  {Thornley} M.~D.,  {Regan} M.~W.,  {Wong} T.,  {Sheth} K.,
  {Vogel} S.~N.,  {Blitz} L.,   {Bock} D.~C.-J.,  2003, \mn@doi [\apjs]
  {10.1086/346076}, \href {http://adsabs.harvard.edu/abs/2003ApJS..145..259H}
  {145, 259}

\bibitem[\protect\citeauthoryear{{Helou}, {Soifer}  \&
  {Rowan-Robinson}}{{Helou} et~al.}{1985}]{Helou+1985}
{Helou} G.,  {Soifer} B.~T.,   {Rowan-Robinson} M.,  1985, \mn@doi [\apjl]
  {10.1086/184556}, \href {http://adsabs.harvard.edu/abs/1985ApJ...298L...7H}
  {298, L7}

\bibitem[\protect\citeauthoryear{{Helou} et~al.,}{{Helou}
  et~al.}{2004}]{Helou+2004}
{Helou} G.,  et~al., 2004, \mn@doi [The Astrophysical Journals]
  {10.1086/422640}, \href {http://adsabs.harvard.edu/abs/2004ApJS..154..253H}
  {154, 253}

\bibitem[\protect\citeauthoryear{{Hinz} et~al.,}{{Hinz}
  et~al.}{2004}]{Hinz+2004}
{Hinz} J.~L.,  et~al., 2004, \mn@doi [The Astrophysical Journal Supplement
  Series] {10.1086/422558}, \href
  {https://ui.adsabs.harvard.edu/abs/2004ApJS..154..259H} {154, 259}

\bibitem[\protect\citeauthoryear{{Honig} \& {Reid}}{{Honig} \&
  {Reid}}{2015}]{Honig+2015}
{Honig} Z.~N.,  {Reid} M.~J.,  2015, \mn@doi [\apj]
  {10.1088/0004-637X/800/1/53}, \href
  {http://adsabs.harvard.edu/abs/2015ApJ...800...53H} {800, 53}

\bibitem[\protect\citeauthoryear{Israel et~al.,}{Israel
  et~al.}{1984}]{Israel+1984}
Israel F.,  et~al., 1984, Astronomy and Astrophysics, 134, 396

\bibitem[\protect\citeauthoryear{{Jog}}{{Jog}}{1992}]{Jog+1992}
{Jog} C.~J.,  1992, \mn@doi [\apj] {10.1086/171289}, \href
  {https://ui.adsabs.harvard.edu/abs/1992ApJ...390..378J} {390, 378}

\bibitem[\protect\citeauthoryear{{Jog} \& {Solomon}}{{Jog} \&
  {Solomon}}{1984}]{Jog+1984}
{Jog} C.~J.,  {Solomon} P.~M.,  1984, \mn@doi [\apj] {10.1086/161598}, \href
  {https://ui.adsabs.harvard.edu/abs/1984ApJ...276..127J} {276, 127}

\bibitem[\protect\citeauthoryear{{Kennicutt} Jr. et~al.,}{{Kennicutt}
  et~al.}{2003}]{Kennicutt+2003}
{Kennicutt} Jr. R.~C.,  et~al., 2003, \mn@doi [\pasp] {10.1086/376941}, \href
  {http://adsabs.harvard.edu/abs/2003PASP..115..928K} {115, 928}

\bibitem[\protect\citeauthoryear{{Kennicutt} et~al.,}{{Kennicutt}
  et~al.}{2011}]{Kennicutt+2011}
{Kennicutt} R.~C.,  et~al., 2011, \mn@doi [\pasp] {10.1086/663818}, \href
  {http://adsabs.harvard.edu/abs/2011PASP..123.1347K} {123, 1347}

\bibitem[\protect\citeauthoryear{{Leroy}, {Walter}, {Brinks}, {Bigiel}, {de
  Blok}, {Madore}  \& {Thornley}}{{Leroy} et~al.}{2008}]{Leroy+2008}
{Leroy} A.~K.,  {Walter} F.,  {Brinks} E.,  {Bigiel} F.,  {de Blok} W.~J.~G.,
  {Madore} B.,   {Thornley} M.~D.,  2008, \mn@doi [\aj]
  {10.1088/0004-6256/136/6/2782}, \href
  {http://adsabs.harvard.edu/abs/2008AJ....136.2782L} {136, 2782}

\bibitem[\protect\citeauthoryear{{Leroy} et~al.,}{{Leroy}
  et~al.}{2009}]{Leroy+2009}
{Leroy} A.~K.,  et~al., 2009, \mn@doi [\aj] {10.1088/0004-6256/137/6/4670},
  \href {http://adsabs.harvard.edu/abs/2009AJ....137.4670L} {137, 4670}

\bibitem[\protect\citeauthoryear{{Leroy} et~al.,}{{Leroy}
  et~al.}{2012}]{Leroy+2012}
{Leroy} A.~K.,  et~al., 2012, \mn@doi [\aj] {10.1088/0004-6256/144/1/3}, \href
  {http://adsabs.harvard.edu/abs/2012AJ....144....3L} {144, 3}

\bibitem[\protect\citeauthoryear{{Mueller} \& {Arnett}}{{Mueller} \&
  {Arnett}}{1976}]{Mueller+1976}
{Mueller} M.~W.,  {Arnett} W.~D.,  1976, \mn@doi [\apj] {10.1086/154873}, \href
  {http://adsabs.harvard.edu/abs/1976ApJ...210..670M} {210, 670}

\bibitem[\protect\citeauthoryear{{Muraoka} et~al.,}{{Muraoka}
  et~al.}{2016}]{Muraoka+2016}
{Muraoka} K.,  et~al., 2016, \mn@doi [\pasj] {10.1093/pasj/psv134}, \href
  {http://adsabs.harvard.edu/abs/2016PASJ...68...18M} {68, 18}

\bibitem[\protect\citeauthoryear{Murphy, Helou, Kenney, Armus  \& Braun}{Murphy
  et~al.}{2008}]{Murphy+2008}
Murphy E.,  Helou G.,  Kenney J.,  Armus L.,   Braun R.,  2008, The
  Astrophysical Journal, 678, 828

\bibitem[\protect\citeauthoryear{{Patra}, {Chengalur}, {Karachentsev}, {Kaisin}
   \& {Begum}}{{Patra} et~al.}{2016}]{Patra+2016}
{Patra} N.~N.,  {Chengalur} J.~N.,  {Karachentsev} I.~D.,  {Kaisin} S.~S.,
  {Begum} A.,  2016, \mn@doi [\mnras] {10.1093/mnras/stv2789}, \href
  {https://ui.adsabs.harvard.edu/abs/2016MNRAS.456.2467P} {456, 2467}

\bibitem[\protect\citeauthoryear{{Poglitsch} et~al.,}{{Poglitsch}
  et~al.}{2010}]{Poglitsch+2010}
{Poglitsch} A.,  et~al., 2010, \mn@doi [\aap] {10.1051/0004-6361/201014535},
  \href {http://adsabs.harvard.edu/abs/2010A%26A...518L...2P} {518, L2}

\bibitem[\protect\citeauthoryear{{Rieke} et~al.,}{{Rieke}
  et~al.}{2004}]{Rieke+2004}
{Rieke} G.~H.,  et~al., 2004, \mn@doi [The Astrophysical Journals]
  {10.1086/422717}, \href {http://adsabs.harvard.edu/abs/2004ApJS..154...25R}
  {154, 25}

\bibitem[\protect\citeauthoryear{Sakamoto, Hasegawa, Hayashi, Handa  \&
  Oka}{Sakamoto et~al.}{1995}]{Sakamoto+1995}
Sakamoto S.,  Hasegawa T.,  Hayashi M.,  Handa T.,   Oka T.,  1995, The
  Astrophysical Journal Supplement Series, 100, 125

\bibitem[\protect\citeauthoryear{{Sandstrom} et~al.,}{{Sandstrom}
  et~al.}{2013}]{Sandstrom+2013}
{Sandstrom} K.~M.,  et~al., 2013, \mn@doi [\apj] {10.1088/0004-637X/777/1/5},
  \href {http://adsabs.harvard.edu/abs/2013ApJ...777....5S} {777, 5}

\bibitem[\protect\citeauthoryear{{Sellwood} \& {Carlberg}}{{Sellwood} \&
  {Carlberg}}{1984}]{Sellwood+1984}
{Sellwood} J.~A.,  {Carlberg} R.~G.,  1984, \mn@doi [\apj] {10.1086/162176},
  \href {https://ui.adsabs.harvard.edu/abs/1984ApJ...282...61S} {282, 61}

\bibitem[\protect\citeauthoryear{{Trumpler}}{{Trumpler}}{1930}]{Trumpler+1930}
{Trumpler} R.~J.,  1930, \mn@doi [Publications of the Astronomical Society of
  the Pacific] {10.1086/124051}, \href
  {http://adsabs.harvard.edu/abs/1930PASP...42..267T} {42, 267}

\bibitem[\protect\citeauthoryear{{Walter}, {Brinks}, {de Blok}, {Bigiel},
  {Kennicutt}, {Thornley}  \& {Leroy}}{{Walter} et~al.}{2008}]{Walter+2008}
{Walter} F.,  {Brinks} E.,  {de Blok} W.~J.~G.,  {Bigiel} F.,  {Kennicutt} Jr.
  R.~C.,  {Thornley} M.~D.,   {Leroy} A.,  2008, \mn@doi [\aj]
  {10.1088/0004-6256/136/6/2563}, \href
  {http://adsabs.harvard.edu/abs/2008AJ....136.2563W} {136, 2563}

\bibitem[\protect\citeauthoryear{{Warren} et~al.,}{{Warren}
  et~al.}{2012}]{Warren+2012}
{Warren} S.~R.,  et~al., 2012, \mn@doi [\apj] {10.1088/0004-637X/757/1/84},
  \href {https://ui.adsabs.harvard.edu/abs/2012ApJ...757...84W} {757, 84}

\bibitem[\protect\citeauthoryear{{Williams}}{{Williams}}{2005}]{Williams+2005}
{Williams} D.~A.,  2005, in Journal of Physics Conference Series. pp 1--17,
  \mn@doi{10.1088/1742-6596/6/1/001}

\bibitem[\protect\citeauthoryear{{Young} \& {Lo}}{{Young} \&
  {Lo}}{1997}]{Young+1997}
{Young} L.~M.,  {Lo} K.~Y.,  1997, \mn@doi [\apj] {10.1086/304909}, \href
  {https://ui.adsabs.harvard.edu/abs/1997ApJ...490..710Y} {490, 710}

\bibitem[\protect\citeauthoryear{{Young}, {van Zee}, {Lo}, {Dohm-Palmer}  \&
  {Beierle}}{{Young} et~al.}{2003}]{Young+2003}
{Young} L.~M.,  {van Zee} L.,  {Lo} K.~Y.,  {Dohm-Palmer} R.~C.,   {Beierle}
  M.~E.,  2003, \mn@doi [\apj] {10.1086/375581}, \href
  {https://ui.adsabs.harvard.edu/abs/2003ApJ...592..111Y} {592, 111}

\makeatother
\end{thebibliography}


\appendix

\section{Rank correlation coefficients}

\begin{table*}
\centering
\normalsize
\caption{\label{corr_irHI} \normalsize \textbf{Spearman's ($\rho$)} and \textbf{Kendall's ($\tau$)} rank correlation coefficients among the gas column densities and dust IR intensities for NGC 3184 and NGC 7793.}
\begin{center}
\begin{tabular}{c|c|cc|cc|cc|cc|cc}
    \hline
    \hline
Galaxy & \multirow{2}{*}{\diagbox{Gas}{Dust}} & \multicolumn{2}{|c|}{8 $\mu$m} & \multicolumn{2}{|c|}{24 $\mu$m} & \multicolumn{2}{|c|}{70 $\mu$m} & \multicolumn{2}{|c|}{100 $\mu$m} & \multicolumn{2}{|c|}{160 $\mu$m}\\
   & & $\rho$ & $\tau$ & $\rho$ & $\tau$  & $\rho$ & $\tau$ &  $\rho$ & $\tau$ & $\rho$ & $\tau$ \\
    \hline
   & Warm N(HI) & 0.696 & 0.507 & 0.518 & 0.368 & 0.558 & 0.390 & 0.574 & 0.409 & 0.590 & 0.422 \\
   & Cold N(HI) & -0.059 & 0.015 & -0.395 & -0.273 & -0.410 & -0.280 & -0.474 & -0.312 & -0.479 & -0.305 \\
  NGC 3184 & Total N(HI) & 0.696 & 0.507  & 0.518 & 0.368  & 0.558 & 0.390 & 0.574 & 0.409 & 0.590 & 0.422 \\
  &  N(H$_2$) & 0.713 & 0.517 & 0.750 & 0.551 & 0.689 & 0.493 & 0.735 & 0.534 & 0.709 & 0.516 \\
   & Total N(H) & 0.817 & 0.623 & 0.785 & 0.594 & 0.721 & 0.528 & 0.803 & 0.607 & 0.777 & 0.584 \\
    \hline
   & Warm N(HI) & 0.205 & 0.133 & 0.344 & 0.229 & 0.412 & 0.279  & 0.513 & 0.352  & 0.622 & 0.440 \\
  NGC 7793 & Cold N(HI) & 0.072 & 0.079 & 0.243 & 0.189 & 0.313 & 0.223 & 0.423 & 0.303 & 0.597 & 0.440 \\
   & Total N(HI) & 0.236 & 0.157 & 0.409 & 0.275 & 0.492 & 0.337 & 0.590 & 0.411 & 0.722 & 0.529 \\
    \hline
    \hline
\end{tabular}
\end{center}
\end{table*}

\begin{center}
\begin{table*}
\centering
\normalsize
\caption{\label{corr_mfir} \normalsize \textbf{Spearman's ($\rho$}) and \textbf{Kendall's ($\tau$)} rank correlation coefficients among the mid-IR and far-IR dust intensities for NGC 3184 and NGC 7793.}
\vspace{0.5cm}
\begin{tabular}{c|cc|cc}
\hline
\hline
\multirow{2}{*}{\diagbox{$\lambda$ - IR}{Galaxy}} & \multicolumn{2}{|c|}{NGC 3184} & \multicolumn{2}{|c|}{NGC 7793} \\
 & $\rho$ & $\tau$ & $\rho$ & $\tau$ \\
\hline
8 vs. 24 $\mu$m  & 0.972  & 0.870 & 0.973  & 0.864  \\
8 vs. 70 $\mu$m  &   0.869  & 0.690 &   0.953  & 0.817   \\
8 vs. 100 $\mu$m  &   0.955  & 0.826  &   0.973  & 0.862 \\
8 vs. 160 $\mu$m  &   0.955  & 0.825 &   0.980 & 0.879    \\
24 vs. 70 $\mu$m  &   0.859  & 0.685 &   0.941  & 0.813  \\
24 vs. 100 $\mu$m  &   0.934  & 0.799 &   0.979 &  0.879  \\
24 vs. 160 $\mu$m  &   0.947  & 0.816 &   0.967  & 0.848 \\
70 vs. 100 $\mu$m  &   0.616  & 0.449 &   0.818  & 0.657  \\
70 vs. 160 $\mu$m  &   0.579  & 0.425 &   0.728  & 0.569   \\
100 vs. 160 $\mu$m  &   0.776  & 0.605 &   0.798  & 0.641 \\
\hline
\hline
\end{tabular}
\end{table*}
\end{center}

\bsp	
\label{lastpage}
\end{document}